\documentclass{aa}  
\usepackage{graphicx}
\usepackage{txfonts}
\usepackage{color}
\usepackage{wasysym}
\usepackage{natbib}
\begin{document}
    \title{The essential elements of dust evolution: \\ 
    a-C(:H) nanoparticle sub-structures and  photo-fragmentation}
       \titlerunning{The essential elements of dust evolution: a-C(:H) nanoparticles}
        \subtitle{}

    \author{A.P. Jones\inst{1}
    \and
    N. Ysard\inst{2} 
    }
    
     \institute{Universit\'e Paris-Saclay, CNRS,  Institut d'Astrophysique Spatiale, 91405, Orsay, France.\\
               \email{anthony.jones@universite-paris-saclay.fr} 
            \and
                   IRAP, Universit\'e de Toulouse, CNRS, UPS, 9Av. du Colonel Roche, BP 44346, 31028 Toulouse, cedex4, France\\
              }

    \date{Received {\em wisdom}: accepted {\em dogmas}}

   \abstract
{Hydrogenated amorphous carbon materials, a-C(:H), are heterogeneous structures consisting of carbon atoms in different hybridisation states and bonding configurations and are thought to constitute a significant and observationally important fraction of the interstellar dust material.  The stability of interstellar a-C(:H) nanoparticles against photo-thermo-dissociation and Coulomb fragmentation needs to take their intrinsic heterogeneity into account.}
{This work aims to characterise semi-conducting a-C(:H) nanoparticle structures and, in particular, their property-characterising aromatic domain size distribution and so predict how they will behave in intense UV radiation fields that can fragment them through dissociative and charge effects as a result of carbon-carbon bond-breaking. }
{Using a statistical approach we determine the typical sizes of the aromatic domains, their size distribution, how they are network-bonded, and where they are to be found within the structure. We consider the effects of thermal excitation, photo-dissociation and charging of a-C(:H) nanoparticles, and the products of their fragmentation.}
{The derived UV photon-induced fragmentation lifetimes for nanometre-sized a-C(:H) nanoparticles, with radii $\sim 0.4-0.5$nm radius and containing $\sim 40-60$ carbon atoms, are of the order of $10^6-10^7$yr in the diffuse interstellar medium and likely $10^2-10^4$ times shorter in photodissociation regions, depending on the local radiation field intensity. Grains larger than this are stable against photodissociation. In H{\footnotesize II} regions only a-C(:H) nanoparticles with radii greater than 0.7nm ($\gtrsim 150$ carbon atoms) are likely to survive.}
{The photon-driven fragmentation of sub-nanometre a-C(:H) particles was determined to be important in the diffuse interstellar medium and also in high excitation regions, such as photodissociation and H{\footnotesize II} regions. However, in these same regions Coulomb fragmentation is unlikely to be an important dust destruction process.}

   \maketitle

\nolinenumbers 

\section{Introduction}
\label{sect_intro}

It has recently been shown that the carbonaceous nanoparticle abundance in Photo-Dissociation Regions (PDRs) is sensitive to the local physical conditions, especially to the hardness and intensity of the interstellar radiation field and that this has important consequences for the gas \citep{2020A&A...639A.144S,2021A&A...649A.148S,2022A&A...666A..49S}. However, the details of the effects working upon the nanoparticles in these regions have yet to be fully understood.  For example it is not yet clear whether the observed nanoparticle depletion in PDRs \citep[e.g.][]{2012A&A...541A..19A,2008A&A...491..797C,2022A&A...666A..49S} is due to thermal, photo-dissociative or charge effects, or a combination of these. The recent analysis and interpretation of James Webb Space Telescope (JWST) Orion Bar NIRCAM and MRS spectroscopic and photometric data \citep{2024A&A...685A..76E} reveals that, in this rather extreme PDR ($G_0 \sim 10^4$), the dust is significantly processed. This work indicates that the shape of the nanoparticle size distribution, in the rather extreme case of Orion, appears to be similar to that of the diffuse InterStellar Medium (ISM) but that the mass associated with the carbonaceous nanoparticles  is reduced by a factor of 15 with respect to the diffuse ISM. In their detailed analysis of the NIRSpec and MRS spectroscopic observations \cite{2024A&A...685A..76E} elucidated variations in the degree of hydrogenation of the carbonaceous nanoparticles, with the least hydrogenated closest to the illuminating stars, suggesting significant UV photo-processing, and the most hydrogenated associated with the colder and denser molecular regions generally populated by larger grains.

Ultraviolet (UV) photons play a pivotal role in the processing of grains in the ISM and in circumstellar regions \cite[e.g.][]{2004A&A...423L..33D,2005A&A...432..895D,2014A&A...569A.119A,2015A&A...584A.123A,2015A&A...581A..92J,2016RSOS....360223J,2016RSOS....360224J,2019A&A...569A.100B,2021A&A...649A..84H}, in that UV irradiation can act directly by dissociating CH and CC bonds or collectively through the thermal excitation of the entire particle, which may then undergo dissociative or sublimation losses. In either case chemical bond breaking is involved and leads to mass loss from the particle, and perhaps to its eventual destruction. Grain charging arises from UV photon absorption leading to electron ejection, and from ion (principally proton) and electron collision and sticking
\cite[e.g.][]{2001ApJS..134..263W,2006ApJ...645.1188W,2016MNRAS.459.2751K}. If a sufficiently high charge state results,  fragmentation due to strong Coulomb repulsion can occur, that is, to a so-called Coulomb explosion. All of these processes place limits on the lifetime of small grains in PDRs and in the diffuse ISM in general. It is these same processes that also regulate the heating and ionisation state of the interstellar gas through the emission of energetic photo-electrons following UV photon absorption. 

Previous work on the Coulomb explosion of grains as a result of catastrophic charging assumed that the grains were homogeneous and that only under special circumstances can this destructive process be important. \cite[e.g.][]{2006ApJ...645.1188W}. If we relax this assumption and allow that most carbonaceous grains in the ISM are amorphous, semi-conducting nanoparticles, and therefore heterogeneous \citep[e.g.,][]{1990MNRAS.247..305J,2012A&A...542A..98J,2016RSOS....360221J,2013A&A...558A..62J}, then Coulomb effects may perhaps have been underestimated. This can arise because the inherent grain constituent binding energies may be less than in homogeneous bulk materials. However, in this case we should probably speak more in terms of Coulomb induced fragmentation rather than Coulomb explosion because the process is most likely to be incremental rather than catastrophic. 

A wealth of earlier work indicates that interstellar carbonaceous nanograins may be an unstable dust population \citep{1994ApJ...420..307J,1995Ap&SS.224..417A,1996A&A...305..602A,1996A&A...305..616A,1999ApJ...512..500J,2010A&A...510A..36M,2010A&A...510A..37M,2011A&A...526A..52M,2011A&A...530A..44J,2012A&A...542A..98J,2013A&A...558A..62J,Faraday_Disc_paper_2014,2014A&A...570A..32B,2015A&A...581A..92J,2016RSOS....360221J,2016RSOS....360223J,2016A&A...588A..43J,2017A&A...602A..46J,2019A&A...627A..38J,2020A&A...639A.144S,2021A&A...649A.148S,2022A&A...666A..49S,2024A&A...685A..76E}. 
In their study of the destruction of polycyclic aromatic hydrocarbons (PAHs) in supernova (SN) shock waves in the Milky Way \cite{2010A&A...510A..36M} derived lifetimes of the order of a few $\times10^8$yr for PAHs with $50-200$ carbon atoms. 
Re-evaluations of the carbonaceous dust destruction by \cite{2011A&A...530A..44J} found lifetimes of $\sim 2.6 \pm 2.4 \times 10^8$yr and, in their more detailed study, \cite{2014A&A...570A..32B} derived lifetimes of $\sim 0.62 \pm 0.56 \times 10^8$yr. Clearly on galactic scales the uncertainties are large \cite[e.g. see][]{2011A&A...530A..44J}.
Early studies specifically concerning carbon dust survival in the diffuse ISM \citep{1994ApJ...420..307J,1995Ap&SS.224..417A,1996A&A...305..602A,1996A&A...305..616A} showed that PAHs with less than $30-40$  carbon atoms are readily dissociated in the diffuse ISM but for those with more than 50 carbon atoms the photo-destruction timescale is more than $10^9$yr. Thus, interstellar sub-nanometre carbonaceous particles would appear to be in a state of constant flux due to the competing processes of destruction, primarily driven by UV photons and including charge induced processes, and re-formation by the collisional fragmentation of larger carbonaceous grains, as proposed by \cite{2022A&A...666A..49S}. This implies that, as in the case of PDRs, the net destruction and/or lack of reformation of the product carbonaceous nanograins dominates \citep{2008A&A...491..797C,2012A&A...541A..19A,2020A&A...639A.144S,2021A&A...649A.148S,2022A&A...666A..49S,2024A&A...685A..76E}.  Thus, there must be environments where these processes are so rapid that they completely deplete the stock of large carbonaceous grains and therefore, ultimately, of all carbonaceous dust. Such regions would obviously show no carbonaceous dust emission bands in the infrared (IR), which may be the case in the interstellar clouds local to the Solar System \citep{2009SSRv..143..311S} and also in some early type and dwarf galaxies \cite[e.g.][]{2008ApJ...672..214G,2018ARA&A..56..673G}. 

This work adopts the THEMIS (The Heterogeneous dust Evolution Model for Interstellar Solids)\footnote{See  http://www.ias.u-psud.fr/themis/} view of interstellar dust \citep[see][]{2013A&A...558A..62J,2014A&A...565L...9K,2015A&A...577A.110Y,2017A&A...602A..46J,2024A&A...684A..34Y}. Here we are most interested in the aromatic rich THEMIS hydrogenated amorphous carbon, a-C(:H), nanoparticles  with radii $\lesssim 5$nm and comprising $\sim 20$\% of the total dust mass, which also contain a significant fraction of aliphatic and olefinic carbon and a residual hydrogen atom fraction of the order of 5\%. The form and structure of these particles was illustrated by \cite{2012ApJ...761...35M} in their proposal for fullerene formation around planetary nebul\ae\ and also shown schematically by \cite{2015A&A...581A..92J}. 

A key and critical component in our consideration of a-C(:H) nanoparticle stability in the ISM is a knowledge of the aromatic domain size distribution in bulk a-C(:H) materials and within a-C(:H) nanoparticles. However, these size distributions are almost certainly not the same because of the structural constraints imposed by finite particle sizes, which are much less restrictive in the case of bulk materials. For bulk carbonaceous materials the aromatic domain size distribution was extensively discussed by \cite{1986AdPhy..35..317R}, \cite{1987PhRvB..35.2946R} and \cite{1988PMagL..57..143R}. We therefore begin with a discussion and a statistical description of the critical structures within a-C(:H) bulk materials and nanoparticles. This is followed by an exploration of the consequences of the molecular structure of nanoparticles within the context of their evolution in a range of interstellar media. 

This paper studies two key aspects of a-C(:H) nanoparticle evolution in the ISM. Firstly, for those wishing to understand the chemical and structural makeup of these materials, Section \ref{sect_aCH_charac} gives an overview of the a-C(:H) structural properties and 
Section \ref{sect_aromatics_network} describes the chemical bonding of the  aromatic domain sub-component within a-C(:H) networks.  
Secondly, Section \ref{sect_photofrag} presents a UV photon-driven fragmentation model and determines a-C(:H) nanoparticlke lifetimes in the ISM and Section \ref{sect_DCF} considers the role of Coulomb fragmentation.  
Finally, Section \ref{sect_conclusions} concludes and summarises this work.

\section{Hydrogenated amorphous carbon, a-C(:H)}
\label{sect_aCH_charac}

Amorphous carbons solids, from richly to poorly hydrogenated, a-C:H to a-C, respectively, encompass a wide compositional and structural range, generically indicated by the a-C(:H) descriptor. 
These materials are semiconducting, spanning the H-rich ($60 \gtrsim$ at.\,\% H $\gtrsim 15$), aliphatic-rich and wide band gap (2.7\,eV $\gtrsim E_{\rm g} \gtrsim 1$\,eV) a-C:H solids at one end to the H-poor ($15 \gtrsim$ at.\,\% H $\gtrsim 2$), aromatic-rich and narrow band gap (1\,eV $\gtrsim E_{\rm g} \gtrsim -0.1$\,eV) a-C at the other extreme  \citep[e.g.][]{1986AdPhy..35..317R,1988PMagL..57..143R,1987PhRvB..35.2946R,2001PSSAR.186.1521R,2002MatSciEng..37..129R,2000PhRvB..6114095F,2004PhilTransRSocLondA..362.2477F}. Their structures consist of contiguous 3D networks of chemically bonded carbon and hydrogen atoms with the carbon atoms principally in sp$^3$ and sp$^2$ hybridisations, in aliphatic, olefinic and aromatic configurations. Carbon atoms in sp hybridisation states are also possible but seemingly somewhat rare and are in any event rather unstable \citep{2004PhilTransRSocLondA..362.2477F}. Quite naturally bi-atomic a-C(:H) materials can include heteroatoms of an element $X$, which are indicated by the descriptors a-C(:H):$X$ or a-C(:H)[:$X$], where the square brackets indicate minor concentrations of element $X$ (e.g. $\ll 10$ atomic\,\%). The most common heteroatoms within a-C(:H)[:$X$] materials are O and N atoms \citep[e.g.][]{2013A&A...555A..39J,2016RSOS....360221J,2016RSOS....360223J,2016RSOS....360224J}.

In this work we are most interested in the aromatic domains within a-C(:H) semiconducting solids because they are the principal determinants of their optical properties \citep[e.g.][]{1986AdPhy..35..317R,1987PhRvB..35.2946R}. They are also the centres for activated conduction and therefore the most likely charge-carrying sites within such materials. Further, it is the $\pi-\pi^\star$ transitions ($E_{{\rm h}\nu}  < 5$eV) of these aromatic domains within a-C(:H) nanoparticles that absorb lower energy photons, while aliphatic and olefinic substructures absorb at somewhat higher energies. Here we explore the nature and size distribution of the aromatic domains within a-C(:H) as a function of the particle size.

The aromatic clusters or domains\footnote{Polycyclic aromatic, aliphatic, and olefinic (hydro)carbon structures wherein the aromatic ring systems are linked by mixed aliphatic-olefinic bridges within an amorphous a-C(:H) network exhibiting no long range order beyond that of the aromatic domains.} can be characterised by the number of their constituent carbon atoms, $n_{\rm C}$, number of aromatic six-fold rings, $N_{\rm R}$, and their mean carbon atom coordination number, $m_{\rm C}$ \cite[e.g.][and references therein]{1990MNRAS.247..305J,2012A&A...540A...1J,2012A&A...540A...2J,2012A&A...542A..98J}. The latter quantity is the mean number of carbon atoms directly bonded to a given carbon atom within the 3D network. In bulk a-C(:H) materials $N_{\rm R}$ is a function of the bulk material band gap, $E_{\rm g}$ \citep{1987PhRvB..35.2946R}, 
that is,
\begin{equation}
N_{\rm R} \simeq \left( 5.8 / E_{\rm g}[{\rm eV}] \right)^2.  
\label{eq_NREg}
\end{equation}
The atomic fraction of hydrogen, $X_{\rm H} = N_{\rm H} / (N_{\rm C}+N_{\rm H})$, within the structure
is a characterising parameter that is also directly relatable to $E_{\rm g}$ via $E_{\rm g} \simeq (4.3 \ X_{\rm H})$eV 
\citep[e.g.][]{1986AdPhy..35..317R}. Another measure of the dimensions of the aromatic domains is the aromatic coherence length, $L_{\rm a}$, which is related to the band gap by $E_g[{\rm eV}] \simeq  0.77 / L_{\rm a}[{\rm nm}] $
\cite[e.g.][]{1986AdPhy..35..317R,1991PSSC..21..199R}. Combining these equations we have  
\begin{equation}
L_{\rm a}[{\rm nm}]  \simeq ( 0.77 \, N_{\rm R}^{0.5} ) / 5.8  = 0.13 \, N_{\rm R}^{0.5}. 
\label{eq_La}
\end{equation}
For compact polycyclic aromatic structures $n_{\rm C}$ \citep[see][Appendix A, Table A.1]{2012A&A...540A...1J}  and the radius of the most compact aromatic domain, $a_{\rm R}$ \citep[see][Eq. 4]{2012A&A...540A...2J}, can be expressed as functions of $N_{\rm R}$,    
\begin{equation}
n_{\rm C} = 2 N_{\rm R} + 3.5\surd N_{\rm R} + 0.5,  
\label{eq_nC_NR1}
\end{equation}
\begin{equation}
a_{\rm R} = 0.09 \left[2 N_{\rm R} + \surd N_{\rm R} + 0.5 \right]^{0.5}  {\rm nm}.  
\label{eq_aR_NR}
\end{equation}
The agreement between Eqs. (\ref{eq_La}) and (\ref{eq_aR_NR}) is within a factor of $\simeq 1.25$ for large $N_{\rm R}$ and within a factor of $\simeq 2$ for $N_{\rm R} = 2$, with $a_{\rm R}$ always larger than $L_{\rm a}$. This difference arises because $a_{\rm R}$ is defined on a geometrical basis whereas $L_{\rm a}$ is determined optically. 

The above expressions indicate a clear link between the bulk material band gap and the size of the intrinsic aromatic sub-structures within a-C(:H) materials. The properties of such materials can be described using a random covalent network model (RCN).  A RCN uses an average nearest-neighbour atomic bonding environment to constrain macroscopic 3D networks of carbon and hydrogen atoms. Within such an atomic network the formation of covalent bonds generally increases stability. However, when that network is not a regular lattice the randomly-directed bonds lead to strain energy resulting from distortions in the bond lengths and angles. The optimal network is one that just balances the effect of bond-induced stability against structural strain. The full details of RCNs have been described elsewhere \citep[e.g.][]{1979PhRvL..42.1151P,1980JNS...42...87D,1986AdPhy..35..317R,1990MNRAS.247..305J} and in its extended form \citep[eRCN,][]{2012A&A...540A...1J}.

\section{Aromatic domain network connections in a-C(:H)}
\label{sect_aromatics_network}

How well a given aromatic domain connects within an a-C(:H) network structure is determined by the number of its edge sites that link to other parts of the contiguous network. In the eRCN model \citep{2012A&A...540A...1J} not all edge sites can link because this leads to a highly-strained system that would tend to stabilise by de-linking to form dangling bonds. Such non-linking, dangling edge sites are not allowed in the eRCN model and are therefore assumed to be singly hydrogenated. It is this structure versus strain offset that the RCN and eRCN models take into account in what is, essentially, a statistical description of the nanoscopic structure and short range order within a-C(:H) solids. The aromatic domain connectivity within the RCN can be determined using the average cluster coordination number $m_{\rm cluster} = (3.5 \surd N_{\rm R}+2.5)$, with lower and upper limits of ($N_{\rm R}+6$) and ($2 N_{\rm R}+4$), respectively \citep[see][Appendix A, Table A.1]{2012A&A...540A...1J}.

Particularly important in the case of hydrogen-poor eRCNs is the need to consider the  hydrogen content of the aromatic-linking, mixed bonding aliphatic (sp$^3$) and olefinic (sp$^2$) network substructures in addition to that of the sp$^2$ aromatic domains. The aromatic cluster coordination number per carbon atom, $m_{\rm ar}$, is given by the cluster coordination number, $m_{\rm cluster}$, divided by the number of constituent carbon atoms, $n_{\rm C}$, 
\begin{equation}
m_{\rm ar} = \frac{m_{\rm cluster}}{n_{\rm C}} = \frac{3.5 \surd N_{\rm R}+2.5}{2N_{\rm R}+3.5\surd N_{\rm R}+0.5},   
\end{equation}
where it is assumed that the aromatic domains are compact. From \cite{2012A&A...540A...1J} the atomic fraction of carbon in sp$^2$ aromatic form bonded to a hydrogen atom, $X^2_{\rm CH,ar}$, (i.e. in aromatic CH bonds) is given by 
\begin{equation}
X^2_{\rm CH,ar} =  \eta \, m_{\rm ar} \, f X_{sp^2},  
\end{equation}
where $f$ is the fraction of sp$^2$ carbon atoms in aromatic domains,\footnote{As per \cite{2012A&A...542A..98J} 
$f  =  f_{\rm max} / [ {\rm e}^{(X_{\rm H}-X_{\rm Hc})/\delta} + 1 ]$, 
where the maximum fraction of sp$^2$ carbon atoms that can be incorporated into aromatic clusters $f_{\rm max}= 0.6$ [= 0] at low [high] $X_{\rm H}$, $X_{\rm Hc} = 0.33$ and  $\delta = 0.07$ defines the steepness of the transition between high and low $f$.} $X_{sp^2}$ is the atomic fraction of sp$^2$ C atoms, which can be approximated\footnote{This approximation is derived from Eq. (24) of \cite{2012A&A...540A...1J}, viz. $X_{sp^2} =  (8-13 X_{\rm H})/2(7-0.5[7-b])$, with the substitutions $b=0.5(7-Y_f)$ and $Y_f=(7-2Z)f$, as stated just before Eq. (40). $Z$ is given by Eq. (A.1) of \cite{2012A&A...540A...1J} and for $N_{\rm R} =1-3$ yields $Z=2$ to the nearest integer and the denominator simplifies to $2(7-0.5[7-3f])$.} by 
\begin{equation}
X_{sp^2} =  \frac{(8-13 X_{\rm H})}{2(7-0.5[7-3f])},   
\label{eq_Xsp2}
\end{equation}
and $\eta = [H]/[C] = X_{\rm H}/(1-X_{\rm H})$ is the hydrogen to carbon atom ratio for bulk a-C(:H) materials, which is assumed to be the same for both sp$^2$ and sp$^3$ carbon atoms. Adopting the same hydrogen to carbon atom concentration ratio for olefinic and aliphatic carbon the fraction of hydrogenated aromatic domain edge sites, $\zeta$, is  given by 
\begin{equation}
\zeta  = \frac{X^2_{\rm CH,ar} }{f X_{sp^2}} = \eta \, m_{\rm ar}.  
\end{equation}
$(1 - \zeta)$ is then the fraction of aromatic domain edge sites that link to the network and $( 1 - \zeta) \, m_{\rm cluster}$ the number of network links for a given $N_{\rm R}$. For nanoparticles consisting of a-C(:H) materials in the diffuse ISM, and particularly in PDRs and H{\footnotesize II} regions, the grains are likely to be predominantly H-poor and therefore of a-C composition with properties equivalent to those of bulk materials with $E_{\rm g} \simeq 0.1$\,eV \citep{2013A&A...558A..62J} or even as low as $0.03$eV in the Orion Bar PDR \citep{2024A&A...685A..76E}. Fig. 2 of \cite{2012A&A...542A..98J} shows that a-C(:H) nanoparticles, with radii from 0.4nm to 1nm, equivalent to the $\sim 3-13\mu$m band emitters modelled by \cite{2024A&A...685A..76E}, have actual band gaps ranging from $\sim 2.5$eV to $\sim 0.8$eV, respectively. For nanoparticles in the diffuse ISM with an effective bulk band gap $E_{\rm g} \simeq 0.1$\,eV, $X_{\rm H} \simeq 0.02$ and $X_{sp^2} = 0.88$, we find $\zeta \simeq 0.01$ to 0.02, for  aromatic domains with $N_{\rm R} = 1-10$, $n_{\rm C} = 6-32$ and $m_{\rm cluster} = 6-14$.\footnote{N.B. For $E_{\rm g} \simeq 0.1$\,eV bulk materials Eq. (\ref{eq_NREg}) predicts aromatic domains with $N_{\rm R} \geq 3000$ and radii $\geq 7$nm that would be  significantly larger than the IR band emitting carriers in THEMIS ($a \sim 0.4-3$nm). Hence the need to consider a size-dependent effective band gap of 2.5eV \citep[see][Fig. 2]{2012A&A...542A..98J}, which yields $N_{\rm R} \sim 5$.} The derived values of $\zeta$ are therefore practically independent of size for the most abundant aromatic domains in a-C nanoparticles \citep[i.e. $N_R = 1 - 5$,][]{2012A&A...542A..98J}. 

The overriding conclusion of this section is that, because of the paucity of H atoms in a-C nanoparticles, 
most of the peripheral aromatic domain carbon atoms must be linked into the 3D network at low values of $X_{\rm H}$ in bulk a-C(:H) materials via aliphatic and/or olefinic bonds.

\subsection{Size-dependent cluster network connections}
\label{sect_sized_aromatics_network}

\begin{figure} 
\centering
 \includegraphics[width=10.cm]{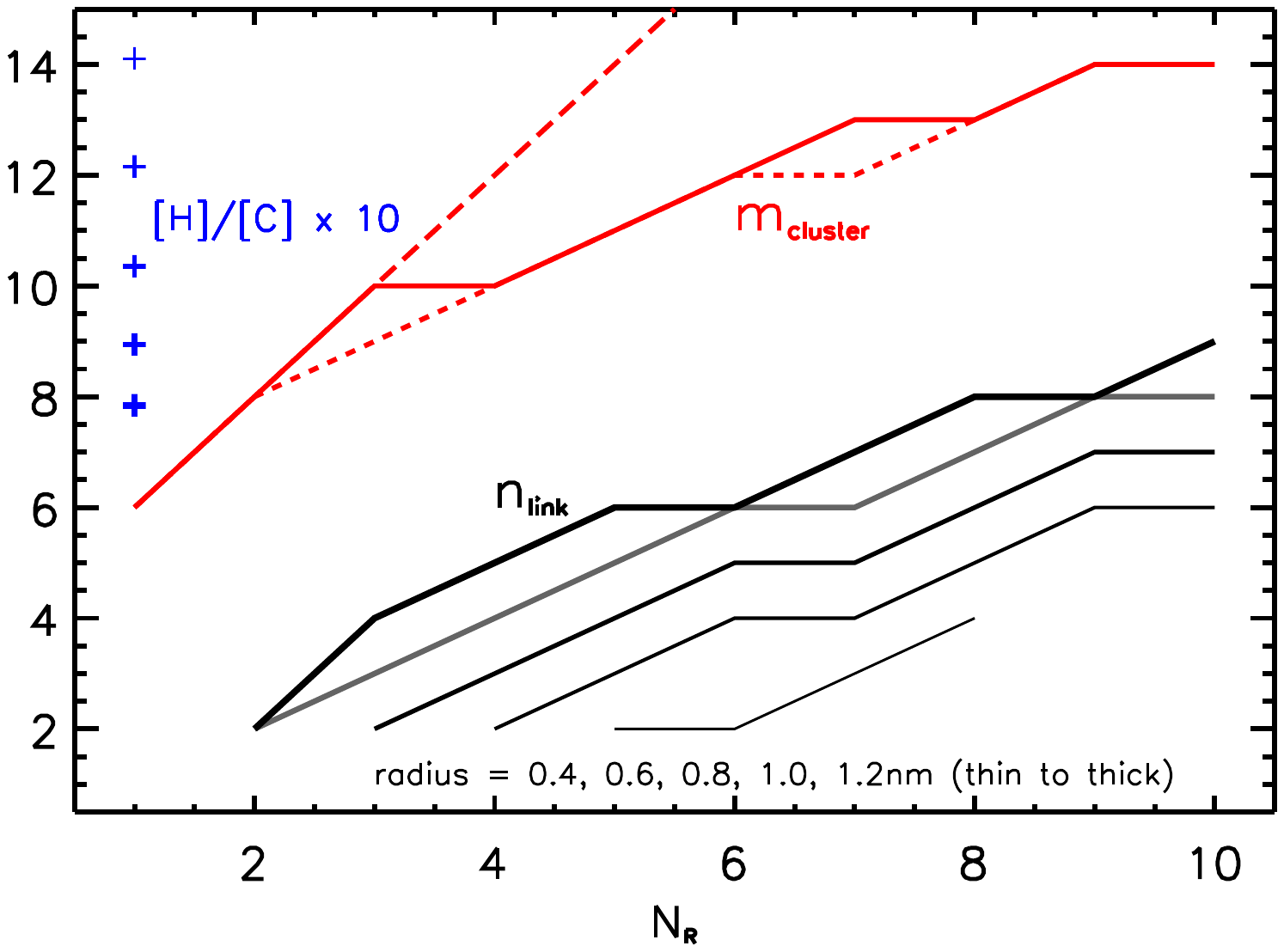}
 \caption{Aromatic clusters coordination numbers, $m_{\rm cluster}$ (red lines, upper and lower limits shown dashed), and number of network linking bonds, $n_{\rm link}$, in a-C(:H) nanoparticles as a function of  $N_{\rm R}$ and radius (0.4, 0.6, 0.8, 1.0, 1.2\,nm, thin to thick black and grey lines, respectively). For each particle radius (same line thickness) the blue cross ordinate indicates the size-dependent hydrogen to carbon atomic ratios, [H]/[C], multiplied by 10.}
\label{fig_links}
\end{figure}

Within the framework of  THEMIS a major fraction of the carbon atoms in the a-C(:H) nanoparticles are found to be in aromatic domains, held together by aliphatic-olefinic bridges. Such network structures are consistent with their inferred stability and survival in the diffuse ISM.  

As deduced above for bulk materials, the aromatic domain connectivity is determined by the number of edge sites that  link within the network. In this section we reconsider this deduction within the framework of finite-sized particles where  particle surfaces (and, in particular, their hydrogenation) and short-range structural order play a critical role. 

For finite sized particles the number of carbon atoms per particle is given in Appendix A of  \cite{2012A&A...542A..98J} as 
\begin{equation}
N_{\rm C} = 2500 \, \left( \frac{a}{\rm 1\,nm} \right)^3 \, \left( \frac{\rho_{\rm a-C:H}}{\rm 1 \,g/cm^3} \right) \ \frac{(1-X_{\rm H})}{(12-11 X_{\rm H})},
\label{eq_NC_sph}
\end{equation}
where  $a$, is the particle radius, the material density $\rho_{\rm a-C:H} = 1.3 + 0.4 \, {\rm exp}(-(E_{\rm g} + 0.2))$ [g\,cm$^{-3}$] \citep{2012A&A...542A..98J}, and for radii $< 0.5$nm, where the particles are shell-like we can calculate the number of carbon atoms based on the surface area, where 
\begin{equation}
N_{\rm C} = \frac{4 \pi a^2}{\pi r_{\rm atom}^2}= 4 \left( \frac{a}{r_{\rm atom}} \right)^2 \ \ \ \ \ \ \ {\rm for} \ a \ < \ 0.5{\rm nm}, 
\label{eq_NC_shell}
\end{equation}
where $r_{\rm atom}$ is the average radius of the constituent carbon atoms (see below). The hydrogen to carbon atom concentration ratio, $\eta$, for these particles is   
\begin{equation}
\eta  = \frac{[H]}{[C]} =  \frac{X_{\rm H}^\prime}{(1-X_{\rm H})},   
\end{equation}
where $X_{\rm H}^\prime$, the H atom fraction\footnote{Strictly, the H atom fraction, $X_{\rm H}$, only applies to the bulk but because the particle surface is also considered here the `fraction' $X_{\rm H}^\prime$ can take values greater than unity.} that takes into account surface passivation by hydrogen atoms \citep{2012A&A...542A..98J}\footnote{The RCN model applies to bulk materials and takes no account of a surface state when applied to finite-sized particles. Thus, outwardly-directed, dangling surface atomic bonds are assumed to be bonded to abundant, monovalent hydrogen atoms as per the eRCN model.}, is
\begin{equation}
X_{\rm H}^\prime = X_{\rm H} + 0.5 (1+X_{\rm H}) \, F_S m_{\rm C},    
\end{equation}
where $F_S$ is the fraction of carbon atoms at the surface 
\begin{equation}
F_S =  \Bigg\{ 1 - \left[ 1 - \left( \frac{2 \, r_{\rm atom}}{a} \right) \right]^3 \Bigg\}.  
\end{equation}
and $r_{\rm atom}$ is the average radius of a carbon atom, defined as 
\begin{equation}
r_{\rm atom} = \left( \ \frac{N_{\rm ar}}{N_{\rm C}} L_{\rm CC}({\rm ar}) \, + \frac{N_{\rm ol}}{N_{\rm C}} L_{\rm CC}({\rm ol}) \, + \frac{N_{\rm al}}{N_{\rm C}} L_{\rm CC}({\rm al}) \ \right),  
\end{equation}
with $N_j/N_{\rm C}$ ($j =$ ar, ol, al) the C atom fraction of aromatic, olefinic and aliphatic C atoms, and $L_{\rm CC}(j)$ the CC bond lengths, taken to be 0.140, 0.134 and 0.154\,nm, respectively.\footnote{The bond lengths are taken from the National Institute of Standards and Technology (NIST) Computational Chemistry Comparison and Benchmark DataBase (CCCBDB) and are those for benzene, cylcohexene and adamantane. https://cccbdb.nist.gov/expbondlengths1x.asp \label{foot_bonds}} In the eRCN model the mean coordination number of the carbon atoms, $m_{\rm C}$, can be expressed as 
\begin{equation}
m_{\rm C} \ \ \ \ \ \ \ \, = \frac{(20-15 X_{\rm H}) + 4 \, [7-2Z] f \, (1-X_{\rm H})}{(7+ [7-2Z] f) \, (1-X_{\rm H})},       
\label{eq_mC}
\end{equation}
where $Z$ is the number of constraints per carbon atom for the cluster, that is $Z = N_{{\rm con},N_{\rm R}} / n_{\rm C}$ where $N_{{\rm con},N_{\rm R}} = (2.5 \, m_{\rm coord} - 3)$ \citep{2012A&A...540A...1J}, and  
\begin{equation}
Z = \left[ \frac{5}{4} (7 \surd \langle{N_{\rm R} \rangle} +5) - 3 \right] \frac{1}{N_{\rm ar}}, 
\end{equation}
here $\langle N_{\rm R} \rangle$ is the average number of six-fold rings per aromatic domain for the nanoparticle composition under consideration. As above, the fraction of hydrogenated aromatic domain edge sites is $\zeta$ and  $( 1 - \zeta) \, m_{\rm cluster}$ the number of aromatic domain edge sites that link to the network for a given 
$N_{\rm R}$.

The small (nanoparticle) a-C(:H) grains in the diffuse ISM, PDRs and H{\footnotesize II} regions, as mentioned at the end of the preceding section, are assumed to have $E_{\rm g} \simeq 0.1$eV,\footnote{In the atomic region of the Orion PDR the a-C(:H) nanoparticle band gap is reduced to 0.03\,eV \citep{2024A&A...685A..76E}.} $f = 0.6$, and $X_{sp^2} = 0.88$, which yields $\eta =$ [H]/[C] $\simeq 1.4$ to 0.8 for nanoparticles with radii from 0.4 to 1.2nm, respectively. 

Within the eRCN framework a given aromatic domain can only be counted as a sub-component of the contiguous a-C(:H) network if, firstly, $n_{\rm C}(N_{\rm R})$ is less than the total number of carbon atoms in the nanoparticle and, secondly, that it is connected to the network by at least two peripheral, linking single bonds.\footnote{These peripheral aromatic bonds must be single and to sp$^3$ or sp$^2$ carbon atom structures such as $\hexagon -$CH$_2-$ or $\hexagon -$CH=, respectively, in order to preserve the aromatic character of the domain within the network.} The aromatic domain coordination number, $m_{\rm cluster}$, and the number of network linking aliphatic CC bonds, $n_{\rm link}$, are shown by the red and black lines, respectively, in Fig. \ref{fig_links}, and given in Table~\ref{table_links}, as a function of the number of aromatic rings per domain, $N_{\rm R}$. The values of $n_{\rm link}$ are size dependent and shown for nanoparticle radii of 0.4, 0.6, 0.8, 1.0, and 1.2\,nm, where the line thickness increases with increasing radius. For each radius the particle size-dependent hydrogen to carbon atomic ratios, [H]/[C]$\times 10$, are indicated by the ordinates of the blue crosses. Fig. \ref{fig_links} shows that single rings are not favoured (i.e. $N_{\rm R} \geqslant 2$), which is because of the higher values of [H]/[C] ($= 1$ for benzene)\footnote{The [H]/[C] ratios for other fully hydrogenated aromatics, with 2, 3, 4, and 7 rings, are 0.8, 0.71, 0.62, and 0.5 for naphthalene, anthracene, pyrene and coronene, respectively.} that they would impose. Equivalently, the weak increase in the cluster coordination number with increasing $N_{\rm R}$ (see Table \ref{table_links}) significantly reduces the network linking per carbon atom. With increasing radius increasingly smaller aromatic domains can be accommodated and for $a \gtrsim 1.3$\,nm this encompasses single aromatic rings. Thus, for larger grains and bulk materials benzene-type aromatic rings can occur within the structures. Na\"{i}vely, it would seem that the smallest particles ought to be able to encompass single, benzene-like aromatic domains. However, and perhaps paradoxically, Fig. \ref{fig_links} shows that sub-nanometre radius particles, with $\sim 40-80$ carbon atoms ($a \sim 0.4-0.8$nm), actually tend to favour larger aromatic domains ($N_{\rm R} \sim 4-8$) comprising $\sim 20-30$ carbon atoms. This apparent paradox arises because the RCN model balances the nearest neighbour covalent bonding constraints against strain energy introduced by bond angle and bond length distortions. In conclusion, a significant fraction of the carbon atoms in a-C(:H) nanoparticles must be in aromatic domains, which is consistent with their inferred composition and stability in the diffuse ISM \citep{2013A&A...558A..62J}.  

A comparison of the $m_{\rm cluster}$ and $n_{\rm link}$ behaviour in Fig. \ref{fig_links} indicates that only about half of the peripheral carbon atoms in the aromatic domains are linked to the network via single CC bonds in nanometre-sized particles. The degree of linking increases with increasing radius but is only weakly dependent on $N_{\rm R}$.  Thus, 
the weakest bound domains, with the fewest network links, are the smallest, which has interesting consequences for a-C nanoparticle photo-dissociation in the ISM (see Section \ref{sect_TED}). 

Support for small aromatic domains in a-C(:H) nanoparticles comes from steric considerations, which imply that the aromatic clusters should principally consist of isolated two- and three-ring systems containing about three quarters of the carbon atoms, see \cite{2012A&A...542A..98J} and Figs. 1 to 4 in \cite{2012ApJ...761...35M}. The sub-nanometre, aromatic-aliphatic structures shown in \cite{2012ApJ...761...35M} were energetically minimised to ensure the structural integrity of the cluster by balancing the constraints imposed by bond angle and bond length distortions. The non-aromatic carbon atoms must be in aliphatic-olefinic bridging structures, mostly in the form of short chains with typically $2-6$ carbon atoms: the only possible four carbon atom bridging structures are --CH$_2$--CH$_2$--CH$_2$--CH$_2$--, --CH$_2$--CH=CH--CH$_2$--, or --CH=CH--CH=CH-- \citep{2015A&A...581A..92J}. Note that the terminating bonds in these bridges must be single in order to preserve the aromaticity of the adjacent aromatic domains. Further, the flexibility of the aliphatic-olefinic bridges plays a key role in balancing the energies associated with the strain effects due to bond angle and bond length distortions within sub-nanometre a-C(:H) clusters. The aromatic rings are themselves somewhat strained in these structures and lose some of their planarity \cite[e.g. see Figs. 1 to 4 of][]{2012ApJ...761...35M}.  The spectra of sub-nanometre, H-poor a-C(:H) particles (with $E_{\rm g} = 0.1-0.5$\,eV) show predominantly aromatic CH and aliphatic CH$_n$ bands, which is compatible with the hypothesis of cluster structures consisting of aromatic domains with aliphatic-olefinic bridges \citep{2012ApJ...761...35M,2012A&A...542A..98J} and is also consistent with experimental data \citep{2012A&A...548A..40C}.

\begin{table}
\caption{The number of aromatic cluster network linking bonds, $n_{\rm link}$ as a function of the number of aromatic rings, $N_{\rm R}$, and the cluster coordination number, $m_{\rm cluster}$.}
\begin{center}
\begin{tabular}{ccc}
      &         &        \\[-0.35cm]
\hline
\hline
      &         &        \\[-0.35cm]
 $N_{\rm R}$    &    $m_{\rm cluster}$   &   $n_{\rm link}$      \\[0.05cm]
\hline
               &                    &                \\[-0.35cm]
    2         &        8          &     2-3       \\
    3         &      9-10       &     3-4       \\    
    4         &     10-12      &     2-5       \\   
    5         &     11-14      &     2-6       \\ 
    
    6         &     12-16      &     3-7       \\ 
    7         &     12-18      &     3-7       \\ 
    8         &     13-20      &     4-8       \\ 
    9         &     14-22      &     6-9       \\ 
  10         &     14-24      &     7-9       \\ 
\hline
\hline
           &         &         \\[-0.25cm]
\end{tabular}
\end{center}
\label{table_links}
\end{table}

\subsection{Aromatic domain size distribution}
\label{sect_aromatics_sdist}

In any given a-C(:H) material the maximum aromatic domain size, $N_{\rm R}$(max), is given by that for a bulk material \cite[e.g. Eq.~(\ref{eq_NREg}) and Fig.~23 from][]{1986AdPhy..35..317R}.  For finite-sized particles, such as interstellar grains, $N_{\rm R}$(max) is further constrained by the particle size, $N_{\rm R}(a)$, because the aromatic domains cannot be larger than the largest particle dimension. In fact they need to be smaller than this in order that the particle hangs together, that is, it is contiguous. $N_{\rm R}(a)$ can then be obtained by solving for $N_{\rm R}$ in Eq.~(\ref{eq_aR_NR}). For a-C(:H) particles with radii $a < 1$nm ($N_{\rm C} \lesssim 300$),  $N_{\rm R}$(max) $\lesssim 60$, and for the smallest a-C(:H) nanoparticles in the THEMIS diffuse ISM dust model $a = 0.4$nm ($N_{\rm C} \lesssim 40$) we have $N_{\rm R}$(max) $\lesssim 8$ \citep[see Fig. \ref{fig_links} and ][]{2012A&A...542A..98J}. In THEMIS it is these a-C(:H) nanoparticles that are predicted to be responsible for the IR emission bands and the UV extinction bump in the diffuse ISM \citep{2013A&A...558A..62J,2017A&A...602A..46J}. 

The largest possible aromatic domain, calculated above, is untenable because it is diametric and would isolate the atoms on either side of it from each other and therefore from forming a part of a contiguous, chemically-bonded, random covalent network structure. Hence, a more appropriate value of $N_{\rm R}$(max) would be equivalent to that of an aromatic domain with a radius of about half of that of the particle radius, that is an aromatic domain diameter equal to the particle radius. With this limitation, a-C(:H) particles with radii $a < 1$\,nm (0.4\,nm) would have $N_{\rm R}$(max) $ \lesssim 13$ (2). This is in reasonable agreement with the findings of the previous section, that to fulfil the network linking constraints within a-C(:H) nanoparticles requires $N_{\rm R}$(max) $\sim 4-8$. 

The optical properties of the aromatic domains in bulk a-C(:H) materials manifest at UV and optical wavelengths through their $\sigma-\sigma^\star$ and $\pi$--$\pi^\star$ bands, respectively. They are also evident at longer wavelengths through the low energy wing of the $\pi$--$\pi^\star$ band, which was modelled with a power-law $N_{\rm R}$ distribution 
biased towards the smaller sizes  \citep[see][Eq. B.5]{2012A&A...542A..98J}, 
\begin{equation}
n(N_{\rm R}) = \frac{ N_{\rm R}^{-p} \ n_{\rm C}(N_{\rm R}) }{ \sum n_{\rm C,arom.} },
\label{eq_aromatics_size_dist}
\end{equation}
where $n(N_{\rm R})$ is the relative abundance of aromatic ring systems with $N_{\rm R}$ rings. The power law exponent in this distribution, $p$, varies from 2.5 to 3.55 for the highest to lowest band gap materials, respectively,\footnote{The aromatic cluster power-law, $p$, for bulk a-C(:H) materials is given by $p \simeq 3.53 - 0.19 \times E_{\rm g}^2$ \citep[][Eq. B.9]{2012A&A...542A..98J}} and  $n_{\rm C}(N_{\rm R})$ is the number of carbon atoms in an aromatic cluster with $N_{\rm R}$ rings given by Eq.~(\ref{eq_nC_NR1}). Note that the cluster abundance is normalised by $\sum n_{\rm C,arom.}$, the total number of carbon atoms in aromatic clusters per unit volume. However well this might work for bulk materials \citep[e.g. see][Appendix B]{2012A&A...542A..98J}, this same kind of power-law distribution cannot apply to nanoparticles because the relatively small number of carbon atoms per particle implies, as predicted by the steepness of the power-law, that the aromatic domain size distribution will be dominated by the smallest domains, each with only a few rings, that is $N_{\rm R} = 1$ to 5. This is clearly in contradiction with the above network linking results and is also in conflict with the requirement that the band gap of such particles is dominated by the presence of aromatic domains that are as large as possible for the given particle radius. Thus, for nanoparticles with less than $\simeq 10^3$ carbon atoms a different approach is needed, which begins with the inclusion of the largest possible aromatic domain for the given band gap and particle size \citep{2012A&A...542A..98J}. In this case, it is easy to imagine a particle in which the aromatic domains are effectively populated top-down starting with the largest possible domain and adding, depending on the residual carbon atoms, sequentially-smaller domains until the requisite number of aromatic carbon atoms is satisfied. This would be equivalent to a rather flat aromatic domain number distribution with some domain sizes unoccupied. Thus, a power-law type of aromatic domain distribution is almost certainly not appropriate for a-C(:H) nanoparticles because their cluster-like structures are finite and dominated by critical size-related effects, which do not apply in bulk materials. We therefore do not consider  power law size distributions for the aromatic domains in nanoparticles but assume the top-down approach described above.

\subsection{Aromatic cluster spatial separation}
\label{sect_aromatics_sep}

Of particular interest in the nanophysics of a-C(:H) particle processing and (photo)fragmentation is the geometrical arrangement of and distance between the aromatic domains in these semi-conducting materials. This is important because the aromatic domains within a-C(:H) nanoparticles are the localised charge-carrying sites responsible for their electrical conductivity. These sites are isolated and therefore  charge mobility, or electrical conduction, in a-C(:H) is necessarily excitation-driven. 

a-C(:H) nanoparticles are richer in hydrogen than their bulk counterparts because of the need for surface passivation, in the absence of dangling bonds, and are also of lower effective density, with respect to bulk matter \citep[see][Appendix A]{2012A&A...542A..98J}.  As shown by \cite{2012ApJ...761...35M} a-C(:H) nanoparticles with radii $\simeq 0.7$nm containing about 100 carbon atoms (see their Figs. ~3 and 4) likely exhibit very open cluster or shell-like structures. Closed cage-like structure can probably be maintained down to particles with $N_{\rm C} \simeq 50-70$ C atoms, that would be generically fullerene-like and hydrogenated. However, smaller particles will likely be progressively more open, bowl, tube, or ribbon-like, and planar or quasi-linear structures at the smallest particles size scales. This general hypothesis is supported by the laboratory observations and theoretical work of \cite{2004PhilTransRSocLondA..362.2477F} who point out that the larger sp$^2$ clusters in a-C(:H) are rather three-dimensional, cage-like structures while smaller clusters are probably more chain-like. In view of this the aromatic domains in a typical a-C(:H) nanoparticle structure will be separated by, at most, a distance equivalent to the diameter of the particle, which is of the order of $\approx 1$nm in the above, $N_{\rm C} \approx 100$, example taken from \cite{2012ApJ...761...35M}. In larger particles the structure will tend to take on multiple conjoined or interlocking cage-like forms with each characteristic cluster structure having a diameter typically of the order of $0.35-0.5$nm \citep{wang_etal_2001,2012ApJ...761...35M}. In the case of a conducting doubly-charged or di-cation particle a cage-like structure can be used to estimate a minimum condition for charge-induced fragmentation because the charged domains would be, at their most distant, in diametrically-opposed positions on the cage surface. However, a-C(:H) materials are semiconductors in which the charges are localised within aromatic domains (i.e. conduction is excitation-driven) and it is therefore possible, in a di-cationic or higher charged state, that the charged sub-domains may be located closer than the diameter of the particle. 

The number of C atoms per a-C(:H) nanoparticle was given above in Eqs. (\ref{eq_NC_sph}) and (\ref{eq_NC_shell}) \citep{2012A&A...542A..98J}. For the a-C(:H) nanoparticles in the THEMIS model, with $E_{\rm g} = 0.1$eV ($X_{\rm H} = 0.02$), these parameters yield $\rho_{\rm a-C:H} = 1.5$ g\,cm$^{-3}$ and we have 
\begin{equation}
N_{\rm C} \simeq 300 \ \left( \frac{a^{0.8}}{1\,{\rm nm}} \right)^3,   
\label{eq_NC_per_np_THEMIS}
\end{equation}
where the radius is raised to the power of 0.8 to account for the more open structures of sub-nanometre radius a-C(:H) nanoparticles  \citep[see][Appendix A]{2012A&A...542A..98J}. The fraction of carbon atoms in aromatic domains is given by $f X_{sp^2}/(1-X_{\rm H}) = (0.6 \times 0.88)/(1-0.02) = 0.54$. From the earlier discussion in Section \ref{sect_sized_aromatics_network}, if we attribute aromatic domains starting with the lowest allowed values of $N_{\rm R}$ and summing the number of associated carbon atoms (10, 14, 16, 19, 22, 24, 27, 30, 32, for $N_{\rm R} = 2, 3, 4, $\ldots 10)\footnote{This is equivalent to a flat aromatic domain size distribution, as alluded to in the previous section.} until a fraction of $\sim 0.54$ has been used, then we find that the number of aromatic domains per nanoparticle is $\approx (a{\rm [nm]}/0.38)^2$. The number of aromatic domains per a-C(:H) nanoparticles is given by
\begin{equation}
N_{\rm ar}(N_{\rm R}) = f \frac{N_{\rm C}}{n_{\rm C}(N_{\rm R})},  
\end{equation}
with $N_{\rm R} = 3$ to 5, we find that for radii of 0.4, 0.6, 0.8 and 1\,nm, this equation predicts 1, 3-4, 6-8 and 9-13 aromatic domains, 
respectively. Assuming a quasi-spherical, single-shell, cage-like nanoparticle structure, the distance along the surface and between the centres of quasi-circular aromatic domains is  
\begin{equation}
d_{\rm ar} \simeq  \left( \frac{ 8 \times 0.91 }{ N_{\rm ar}(N_{\rm R}) } \right)^{\frac{1}{2}} \left( \frac{a}{\rm 1nm} \right) \ {\rm nm}, 
\label{eq_ar_separation}
\end{equation}
where the factor 0.91 is the area packing efficiency for circular domains. The chordal distance between the aromatic domains is 
\begin{equation}
d_{\rm chord} = 2 \, a  \, {\rm sin} \left( \frac{d_{\rm ar}}{2 \, a} \right) \ {\rm nm}, 
\label{eq_ar_chord}
\end{equation}
where $a$ is the radius in nanometres, giving chordal distances $\simeq 0.75-0.87$nm between the aromatic domains. However, if the radius of the a-C(:H) particle is larger than some characteristic size then the structure is most likely organised into multiple cage-like sub-domains \citep{2004PhilTransRSocLondA..362.2477F}, with typical cage radii of the order of 0.35\,nm \citep[e.g.][]{wang_etal_2001,2012ApJ...761...35M}. Thus, if we simplistically assume spherical cages, with a 60\,\% packing efficiency,\footnote{This packing efficiency, or volume filling factor, $V_{ff}$, is typical of the hexagonal lattice packing of spheres with $V_{ff} = \pi / (3 \surd{3}) = 0.60$, which is less than the closest possible packing of spheres obtained by hexagonal close packing where $V_{ff} = \pi /  (3 \surd{2}) = 0.74$.} assuming a quasi-spherical particle, the number of cages per nanoparticle is 
\begin{equation}
N_{\rm cage} = \frac{\frac{4}{3} \pi (a/1{\rm nm})^3}{\frac{4}{3} \pi (0.35{\rm nm})^3 / 0.6} = 0.6 \left( \frac{a/1{\rm nm}}{0.35} \right)^3, 
\end{equation}
and 0.4, 0.6, 0.8 and 1nm radius a-C(:H) nanoparticles encompass $\simeq 1$, 3, 7 and 14 sub-structure cages, respectively.

\begin{figure} 
\centering
 \includegraphics[width=9.0cm]{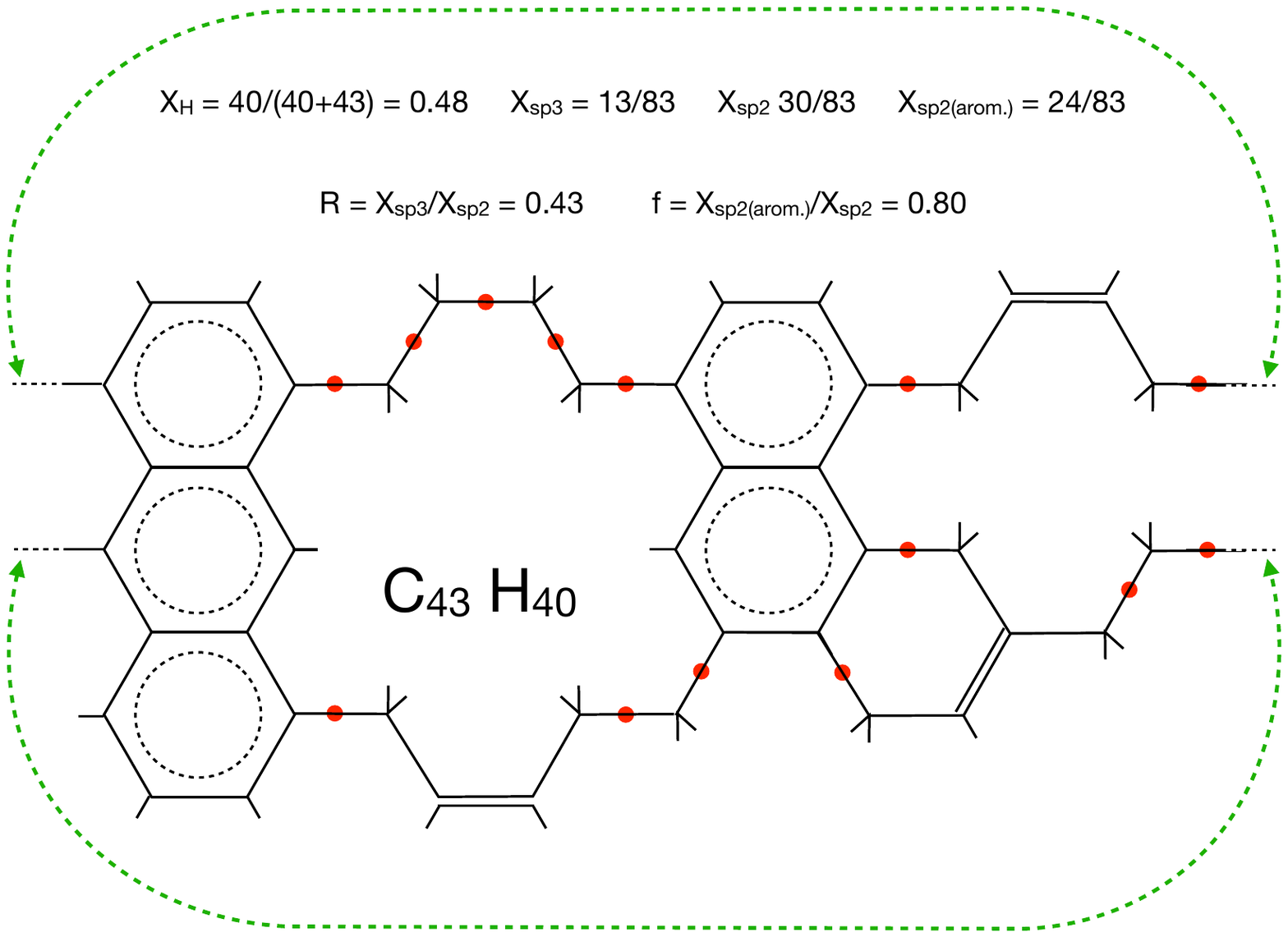}
 \caption{Schematic structure of an idealised, simplified and unfolded, shell-like a-C nanoparticle, C$_{43}$H$_{40}$, which could also be a substructure within a larger a-C(:H) nanoparticle network. The dashed green arrows indicate the common linking bonds that close the shell structure. The red dots indicate the equally probable (i.e. lowest energy) bond cleavage sites. The unadorned, dangling bonds mark the positions of H atoms.}
\label{fig_np_unfolded}
\end{figure}

\section{Photon-driven fragmentation}
\label{sect_photofrag}

There are currently no experimental data\footnote{This is because of the difficulty in forming, maintaining and isolating such nanoparticles under experimental conditions.} or theoretical models to aide us in predicting the fragmentation pathways of the contiguous amorphous structures that are typical of a-C(:H) nanoparticles, as illustrated in Figs. 3 and 4 of \cite{2012ApJ...761...35M}. We will therefore base our modelling of the photon-driven fragmentation of these nanograins on the methods applied to PAHs \citep[][]{1989A&A...213..351L,2010A&A...510A..37M} and adapt them and other theoretical modelling to the case of 3D arophatics, structures of mixed aromatic, olefinic and aliphatic bonding \cite[e.g.][]{2012ApJ...761...35M}.

\begin{table}
\caption{Typical hydrocarbon bond-breaking energies. In the polyatomics the energies are for the single C$-$C bonds  indicated by -$\wr$-.}
\begin{center}
\begin{tabular}{ccc}
      &         &        \\[-0.35cm]
\hline
\hline
      &         &        \\[-0.35cm]
 type    &    bond   &   $\sim$bond energy [eV]  \\[0.05cm]
\hline
               &                    &                \\[-0.35cm]
    allyl-adjacent   &      $-$C-$\wr$-C$-$C=C$-$                 &     3.2       \\  
    aliphatic          &      $-$C-$\wr$-C-$\wr$-C-$\wr$-C$-$                   &     3.8      \\  
    hydrocarbon   &        C$-$H                                     &     4.3       \\  
    allyl                 &       $-$C-$\wr$-C=C-$\wr$-C$-$                   &     4.4       \\  
    aromatic         &     C$-$\hspace{-0.25cm}... \hspace{-0.07cm}C      &     5.4       \\   
    olefinic           &     C$=$C                                        &     6.3       \\ 
    alkynic             &     C$\equiv$C                               &     8.7       \\ 
\hline
\hline
           &         &         \\[-0.25cm]
\end{tabular}
\tablefoot{All values, except those for aromatic, olefinic and alkynic CC bonds (taken from the NIST CCCBDB, see footnote \ref{foot_bonds}), are guided by the work of \cite{2017CTC...1115...45}. Note that the hydrogen atoms associated with the CC bonds are not indicated.}
\end{center}
\label{table_bondEs}
\end{table}

In order to calculate the EUV photon-driven fragmentation of a-C(:H) nanoparticles we adopt the THEMIS size-, composition-, and wavelength-dependent absorption efficiency factors, $Q_{\rm abs}(a,E_{\rm g},\lambda)$, to determine the fraction of incident photons absorbed by a given grain. The absorbed energy, $E_{\rm h \nu}$, leads to the internal excitation (electronic and vibrational) of the grain, which then relaxes by dissociation and/or thermal IR emission. The dissociation can be by the loss of small radical fragments, such as C$_2$ in the case of PAHs \citep{2010A&A...510A..36M}, or by the loss of larger aliphatic/olefinic bridging structures (e.g. C$_n$H$_k$, where $n \simeq 2-6$ and $k \gtrsim n$) and the smaller, more weakly bound aromatic domains \citep[$N_{\rm R} \sim 1-3$,][]{2015A&A...581A..92J}. 

Here we are dealing with the destructive process acting on particles that result in the ejection of aromatic domains and the aliphatic-olefinic carbon bridges that link them. The particles that are considered here are small enough that  they can be considered in terms of shell-like clusters with aromatic domains held together by bridges (carbon chains) of type $-$C$-$C$-$C$-$C$-$, $-$C$-$C$=$C$-$C$-$, $-$C$=$C$-$C$-$C$-$ or $-$C$=$C$-$C$=$C$-$ (e.g. see Fig.~\ref{fig_np_unfolded}, where the positions of the CH bonds are indicated by the dangling bonds). Critical clusters probably contain 50-100 atoms ($a \simeq 0.4 - 0.6$nm). The breaking of several linking bridges and/or the ejection of an aromatic domain from a cluster will most likely lead to the disruption (opening up) of the shell structure into a less stable or more easily dissociable form that will be even more susceptible to photon-driven fragmentation, that is it will have an even shorter lifetime. Thus, and as a reasonable approximation, we will assume that the removal of a single linking bridge (i.e. the breaking of two bonds on the same linear bridge) is the rate-determining step in the disruption of a-C(:H) nanoparticles. 

The interesting paper by \cite{2017CTC...1115...45} studies the thermal decomposition of n-hexane and n-hexene isomers, which are almost exact replicas of the mixed aliphatic-olefinic aromatic domain-linking bridges\footnote{The only difference is that the carbon atoms that terminate the linear chains are bonded to one extra hydrogen atom, otherwise they are identical to the hypothesised aromatic domain linking bridges.} that we hypothesise exist within a-C(:H) nanoparticles. Based upon the findings of \cite{2017CTC...1115...45}, and in particular Fig.~1 of that paper, we can draw some key and perhaps surprising conclusions regarding the mechanism of a-C(:H) nanoparticle disruption. The bond energies in Table~\ref{table_bondEs} \citep[taken from ][]{2017CTC...1115...45} indicate that the allyl-adjacent C$-$C bonds\footnote{The allyl bonding configuration is $-$CH$_2-$CH=CH$-$.} are the weakest and therefore the most likely to be broken by thermal- or photo-dissociative disruption, requiring an energy of the order of $ 2 \times 3.2 = 6.4$\,eV to liberate a mixed aliphatic/olefinic linking bridge. 

Further support for this kind of behaviour comes from the well-known fragmentation pathways for cycloalkene molecular ions in mass spectrometry. For cyclohexene, C$_6$H$_{10}$ (a six-fold ring with one C=C bond, -CH$_2$--CH$_2$--CH=CH--CH$_2$--CH$_2$-), this occurs via a retro-Diels-Alder\footnote{A Diels-Alder reaction involves the addition of an alkene, $^{\rm R_1}_{\rm H}>$C=C$<^{\rm R_2}_{\rm H}$, to a conjugated diene, such as butadiene, CH$_2$=CH--CH=CH$_2$, to form a cycloalkene ring system, -CHR$_1$--CH$_2$--CH=CH--CH$_2$--CHR$_2$-.} reaction, which occurs through the simultaneous breaking of the two C--C bonds symmetrically opposite the C=C bond and a structural re-arrangement to yield, [H$_2$C=CH--HC=CH$_2$]$^{+ \bullet}$, a dienyl radical cation and ethene, H$_2$C=CH$_2$. The fragmentation pathway of cycloalkanes is similar leading to ethene and an alkyl resultant radical cation. 

Interestingly, pure aliphatic (CH$_2$)$_{4-5}$ bridges are more resistant than dehydrogenated mixed aliphatic-olefinic bridges. Thus, UV photon-driven dehydrogenation rather than tending toward increasing stability through olefinisation followed by aromatisation will actually, and on the way to increased aromaticity, lead to a weakening of the structure before it becomes more resistant. 

A close look at Fig.~\ref{fig_np_unfolded} indicates that the possible fragments resulting from the breaking of two C$-$C bonds along a single bridging structure would liberate aliphatic and mixed aliphatic-olefinic chains such as 
(CH$_2$)$_{n(=1-4)}$, 
CH$_2-$CH=CH$-$CH$_2$ (C$_4$H$_6$), and 
CH$_2-$CH=CH$-$CH$_2-$CH$_2$ (C$_5$H$_8$). 
If three CC bonds, in the case of branched bridges, are broken during thermal decomposition then branched species resembling 
CH$_2-$CH=C(CH$_2)-$CH$_2$ (C$_5$H$_7$), and 
CH$_2-$CH=C(CH$_2)-$CH$_2-$CH$_2$ (C$_6$H$_9$) are also possible dissociation products. 
In general all of these break-down fragments can be characterised by the following formula
\begin{equation}
{\rm C}_n  {\rm H}_{2(n-m)}  ({\rm CH})_p, 
\label{eq_generics}
\end{equation}
where $n$ is the number of C atoms in the longest chain in the fragment (excluding branchings), $m$ is the number of double bonds in the longest chain and, in this case, $p$ is the number of CH$_2$ branchings. With the restrictions that $m \leq (n-1)/2$ [for odd $n$], $m \leq n/2$ [for even $n$], and $p \leq (n-2)$. This formula can be adapted for branchings other than by CH$_2$ groups simply by replacing (CH) by the relevant pendant group (less one H atom). However, it is obvious that the energy required to release a fragment increases with the number of branchings, that is,  with increasing $p$, and the probability of their being released must therefore be much lower than for simple linear bridging species. 

Clearly the dissociation products described above may sub-fragment upon ejection or will themselves be subject to photo-dissociation into smaller and simpler secondary dissociation species \cite[e.g. see section 3.3 in][]{2015A&A...581A..92J}. As Fig. \ref{fig_np_unfolded} shows, for an idealised a-C(:H) nanoparticle, the maximum fragments size is in the range C$_4$H$_6$ to C$_5$H$_7$ but most are expected to be somewhat smaller than this. 

It appears that the above UV photon driven a-C(:H) nanoparticle decomposition mechanism could explain the origin of the  
C$_2$H, C$_3$H, l-C$_3$H$^+$, c-C$_3$H$_2$, l-C$_3$H$_2$, and C$_4$H observed in the Horsehead Nebula PDR by \cite{2012A&A...548A..68P,2005A&A...435..885P} and \cite{2015ApJ...800L..33G}. Within the proposed scenario, these species arise from the UV photo-fragmentation of the bridging aliphatic and olefinic (sub-)structures that link the aromatic domains in a-C(:H) nano-particles \citep{2015A&A...581A..92J}.

\subsection{Photo-thermo-dissocation}
\label{sect_PTD}

The effects of photon-driven nanoparticle destruction can be considered using the photo-thermo-dissocation (PTD) mechanism of  \cite{1989A&A...213..351L}. After photon absorption the excited particle cools and after each cooling step the vibrational energy is redistributed and the particle may partially dissociate through the loss of atoms if the residual internal energy is high enough. These authors considered PAH molecule destruction for both hydrogenated and de-hydrogenated molecules, assuming that in the low density ISM they could be dehydrogenated to PAs. In contrast, the THEMIS a-C(:H) nanoparticles are clearly different from PAHs in that they contain an integral and essential aliphatic carbon component of lower C$-$C bond energy ($\sim 3.8$\,eV) than a typical CH bond ($\sim 4.3$\,eV) or aromatic and olefinic CC bonds ($\sim 5.4$ and 6.3\,eV, respectively). Table \ref{table_bondEs} gives a more complete listing of some typical hydrocarbon bond energies. Thus,  the aliphatic CC bonds in a-C(:H) nanoparticles in the diffuse ISM could be photo-fragmented even while remaining hydrogenated. However, as discussed above for allyl configurations, C$-$C bonds two bonds distant from C$=$C bonds are even weaker ($\sim 3.2$\,eV) and will therefore be the first to break. Hence, the PTD approach needs to be modified for THEMIS nanoparticles in order to account for these hydrogens and the weaker aliphatic, allyl adjacent C$-$C bonds. 

The \cite{1989A&A...213..351L} PTD rate constant for the dissociation of PAH molecules, $k_{\rm d,CC}$, follows the Arrhenius equation
\begin{equation}
k_{\rm d,CC} = A_{\rm CC} \ {\rm exp} \left( \frac{ - B_{\rm CC}}{k_{\rm B} \ T } \right)
\label{eq_Arrhenius}
\end{equation}
where the parameters $A_{\rm CC}$ and $B_{\rm CC}$ are those associated with the requisite C$-$C bond breaking that precedes particle fragmentation. For the dominant case of C$_3$ loss from graphite  \cite{1989A&A...213..351L} adopted $A_{\rm CC} = 1.5 \times 10^{18}$\,s$^{-1}$ and an associated bond energy $B_{\rm CC} = 7.97$\,eV. These numbers yield a dissociation rate constant $k_{\rm d,CC} = 0.710$\,s$^{-1}$ and a lifetime of about one hundredth of a second for a nanoparticle with $N_{\rm C} = 50$ at 1000\,K (the temperature attained upon absorption of a 13.6\,eV photon), assuming that one C$-$C bond per C atom needs to be broken for complete destruction. This result indicates that such PAHs ought to be unstable in diffuse ISM given that they absorb of the order of one hard UV photon per year. Clearly the lifetime increases steeply with size and is sensitive to the exact bond breaking energies involved. Hence, and in the absence of dedicated laboratory data, it is not evident how applicable the PTD formalism is to particles with the kinds structures exhibited by theTHEMIS a-C(:H) nanoparticles.  Nevertheless, in the following we adapt this kind of methodology to the THEMIS nanograins.

\subsection{Photon-induced thermal excitation fragmentation}
\label{sect_TED}

In order to determine the loss rate of the absorbed photon energy via the dissociative loss of C$_n$H$_k$ fragments ($n \sim 2-4$, $k \lesssim 2n$) and via IR photon emission we follow and adapt the method of \cite{2010A&A...510A..37M} that was developed for PAHs and described in their Section 4.1. We also make use of the PAH photo-processing study by \cite{2013A&A...552A..15M}. Here we derive a photon-induced thermal excitation fragmentation process applicable to  heterogeneous a-C(:H) nanoparticles, which is similar to that  applied to PAHs. The fundamental mechanisms are the same, that is, UV photon absorption leads to thermal excitation and particle dissociation by partial fragmentation but do differ in the molecular-level details. The salient points of the methods are reproduced here as adapted to the THEMIS a-C(:H) nanoparticles, which are aromatic-rich but also contain significant aliphatic and olefinic components.

The \cite{2010A&A...510A..37M} approach adopted a microcanonical description for PAH molecules. Here we use the THEMIS a-C(:H) heat capacities \citep{2013A&A...558A..62J} to determine the internal temperatures immediately following UV photon absorption and subsequent energy loss via thermal IR photon emission and/or fragmentation. The particle temperature, $T$, can be derived by solving the following equation for $T$
\begin{equation}
E = \frac{4}{3} \pi \, a^3 \int_{T_0}^T C_{\rm V}(T) \ dT
\label{eq_CvT}
\end{equation}
where $T_0$ is the initial temperature (assumed to be 0\,K,), $a$ is the grain radius and $C_{\rm V}(T)$ its heat capacity. The energy $E$ is initially the absorbed photon energy $E_{\rm h \nu}$ or the internal energy $E_{\rm int}$ following some degree of thermal emission and/or dissociative fragment loss.  We will here assume that the dissociation loss channel involves C$_4$H$_k$ bridging fragments, rather than C$_2$ loss as in the \cite{2010A&A...510A..37M} scheme, and that the energy required to remove these fragments is 6.4\,eV as reasoned in Section \ref{sect_photofrag}. Thus, an a-C(:H) nanoparticle can shed excitation energy via bridging link fragment loss ($E_{\rm loss} = 6.4$\,eV) and/or incremental IR photon emission \cite[e.g.][]{1989A&A...213..351L,2010A&A...510A..37M,2013A&A...552A..15M}. 

The dissociative fragmentation rate constant, $k_{\rm diss}$, can be defined by the Arrhenius form of a unimolecular rate \cite[e.g.][]{2010A&A...510A..37M}, that is  
\begin{equation}
k_{\rm diss} = k_0 \ {\rm exp}\left( \frac{ - E_{\rm loss} }{ k_{\rm B} \, T } \right) 
\end{equation}
with $E_{\rm loss} = 6.4$\,eV the binding energy of the dissociated C$_4$H$_k$ fragment, $k_{\rm B}$ the Boltzmann constant, and $T$ the particle internal temperature calculated from Eq.~(\ref{eq_CvT}). 

The rate constant for IR photon emission, $k_{\rm IR}$, was taken to be fixed at $100$ photons\,s$^{-1}$  by \cite{2010A&A...510A..37M}. The later work by  \cite{2013A&A...552A..15M} showed that, for the highly symmetric and rigid PAHs coronene C$_{24}$H$_{12}$ and circumcoronene C$_{66}$H$_{20}$,  $k_{\rm IR}$ is $\simeq 10$ and $\sim 1-10 $ photons\,s$^{-1}$, respectively, for internal energies in the range $7-15$\,eV. The lower internal energy cut-off is used here because only absorbed photon energies in excess of $\sim 7$\,eV will lead to internal energies that exceed the critical fragmentation threshold energy $E_{\rm loss}$ of 6.4\,eV and that can therefore dissociate a-C(:H) nanoparticles. 

With the THEMIS nanoparticles we are dealing with structures that are much less rigid than PAHs, that clearly exhibit  IR emission modes in addition to their aromatic modes (i.e. aliphatic and olefinic CC and CH emission bands), and that, in their much floppier states, it is here postulated that they are likely to be more efficient IR photon emitters. In contrast to \cite{2013A&A...552A..15M}, we consider that it is the particle internal temperature that is the critical parameter, we adopt a temperature-dependent IR photon emission rate $k_{\rm IR}(T) = 0.02 \times T$s$^{-1}$. The pre-temperature factor of 0.02 was chosen so as to give a reasonable comparison with the results of \cite{2013A&A...552A..15M} and yields $k_{\rm IR}(T) = 2$, 10, 20, and 60s$^{-1}$ at 100, 500, 1000, and 3000\,K, respectively. Note that we do not include an explicit size dependence into $k_{\rm IR}$ but leave this to enter implicitly through the derived, size-dependent internal temperatures. 

It seems reasonable to assume that the energy of the emitted IR photons, $E_{\rm IR}$, must also have some dependence on the nanoparticle temperature and we will here assume an equivalence between the particle internal temperature and the wavelength of the emitted IR photons, implicitly assuming that the particle emits as a blackbody. We have therefore temperature-sliced the $E_{\rm IR}$ values as shown in Table \ref{table_EIR}. The energy dependence of $k_{\rm IR}$ was not considered by \cite{2010A&A...510A..37M}.

\begin{table}
\caption{Emitted IR photon energies, E$_{\rm IR}$, and wavelengths, $\lambda_{\rm IR}$, as a function of particle temperature, $T$, and wavelength ranges, $\lambda$.}
\begin{center}
\begin{tabular}{cccc}
      &         &      &      \\[-0.35cm]
\hline
\hline
      &         &     &   \\[-0.35cm]
 $T$ [K]   &    $\lambda$  [$\mu$m]   &   E$_{\rm IR}$ [eV]   &  $\lambda_{\rm IR}$ [$\mu$m]  \\[0.05cm]
\hline
               &                    &              &        \\[-0.35cm]
    $T > 740$\,K                            &        < 5                      &     0.376    &    3.3   \\
    $740 \geqslant T > 526$\,K      &        5-7                      &     0.200    &    6.2   \\
    $526 \geqslant T > 460$\,K      &        7-8                      &     0.161    &    7.7   \\
    $460 \geqslant T > 334$\,K      &        8-11                      &     0.144    &    8.6   \\
    $334 \geqslant T > 305$\,K      &        11-12                      &     0.110    &    11.3   \\
    305 K $> T$      &        > 12                      &     0.098    &    12.7   \\
    
\hline
\hline
           &         &       &      \\[-0.25cm]
\end{tabular}
\end{center}
\label{table_EIR}
\end{table}

We now consider the competition between the channels for energy loss via IR photon emission, $k_{\rm IR}$, and dissociative fragmentation, $k_{\rm diss}$. Following the methodology of \cite{2010A&A...510A..37M} the (un-normalized) nanoparticle fragmentation probability between the emission of the $n^{\rm th}$ and the $(n +1)^{\rm th}$ IR photons can be expressed as 
$p_i = k_{\rm diss}(E_i)/k_{\rm IR}(E_i)$, where $E_i = (E_{\rm h \nu} - i \times E_{\rm IR}(T))$ (in the nomenclature adopted here). 
However, as \cite{2010A&A...510A..37M} pointed out in their Section 4.1 it is difficult to solve the associated equations (their Eqs. 15 to 17) in closed form and therefore requires some simplifying assumptions. 
Hence, and as per \cite{2010A&A...510A..37M}, assuming that $p_i$ in invariant and given by the average probability, $p_{\rm av}$, the total un-normalised probability of dissociation, $P(n_{\rm max})$, is 
 \begin{equation}
P(n_{\rm max}) = ( n_{\rm max} + 1) p_{\rm av} = \frac{ k_0 \ {\rm exp}(E_{\rm loss}/(k_{\rm B} T_{av}))}{ k_{\rm IR} \ / ( n_{\rm max} + 1)},    
\end{equation}
where $n_{\rm max} = 0.2 \, N_{\rm C}$\footnote{Equivalent to the finding of \cite{2010A&A...510A..37M}, expressed on their page 5, that ``$n_{\rm max} = 10$, 20 and 40 for $N_{\rm C} = 50$, 100 and 200 respectively.''} and taking the average temperature to be the geometric mean $T_{av} = ( T(E_{\rm h \nu}) \times T_{n_{\rm max}})^{0.5}$. 
Here $T(E_{\rm h \nu})$ is the initial temperature after UV photon absorption and 
$T_{n_{\rm max}} = T( E_{\rm int} - n_{\rm max} \times E_{\rm IR})$ is that after the emission of $n_{\rm max}$ IR photons. We consider discrete cooling steps until such time as the internal energy has dropped to the minimum IR photon energy (i.e. $E_{\rm int} = 0.1$\,eV, see Table \ref{table_EIR}). Any given cooling step where $E_{\rm int}$ is greater than $E_0$ results in the loss of 6.4\,eV of internal energy from the particle due the bond breaking resulting in the loss of a C$_n$H$_k$ fragment. 
Otherwise each cooling step results it the loss of an energy E$_{\rm IR}(T)$, that is the energy of the temperature appropriate IR photon (see Table \ref{table_EIR}). Only for the smallest particles and for the initial cooling events following energetic UV photon absorption ($E_{\rm h \nu} \simeq 8$eV) can fragment loss occur. For ionising photons ($E \geqslant 13.6$\,eV) this may result in the loss of two fragments in the first two consecutive cooling events because $13.6$\,eV $> 2 \times E_{\rm loss} = 12.8$\,eV. This approach would appear to be physically reasonable given that thermal dissociation is driven by the highest internal temperatures, which occur immediately after UV photon absorption.

\begin{figure} 
\centering
 \includegraphics[width=9.0cm]{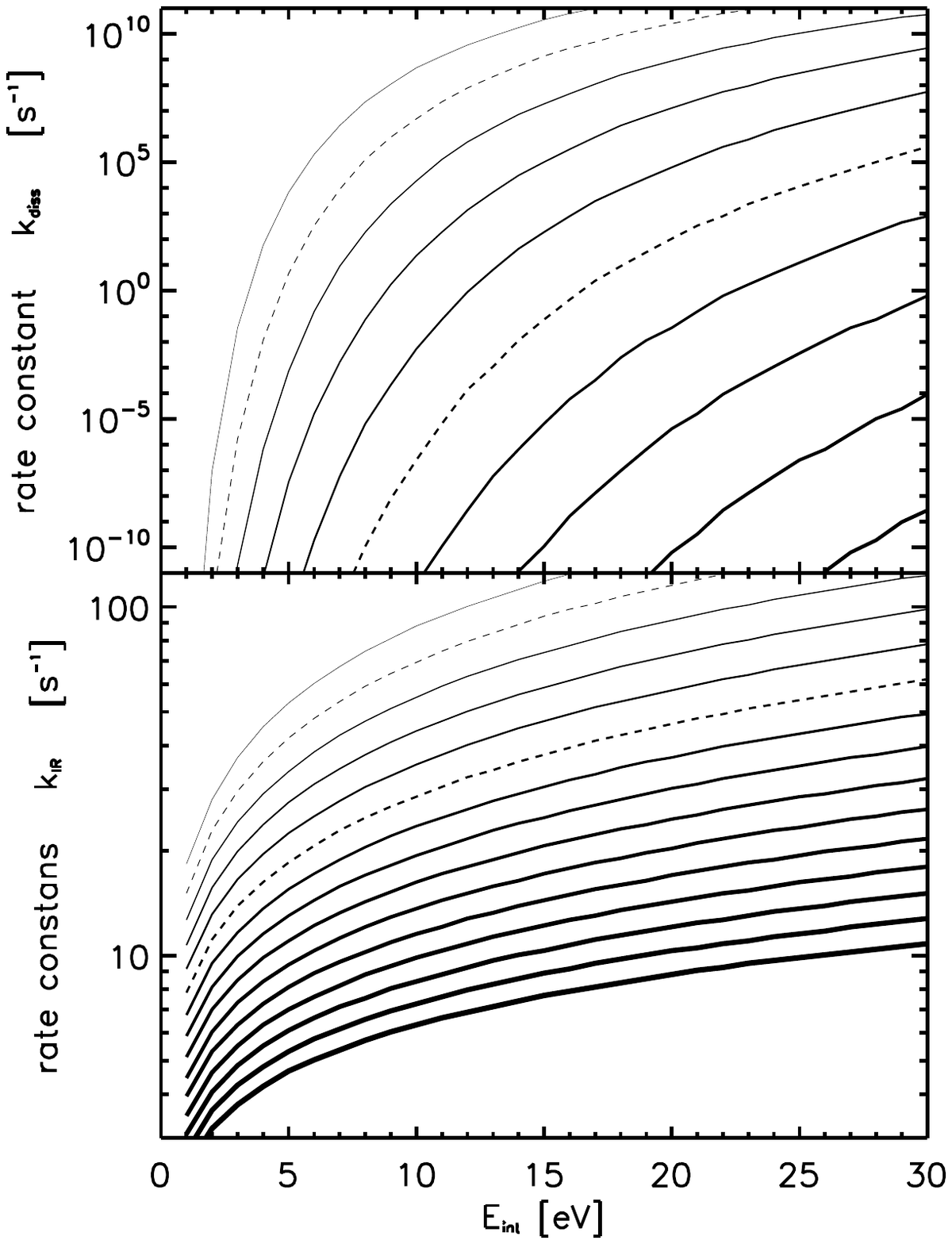}
 \caption{THEMIS a-C(:H) nanoparticle rates constants for dissociation, $k_{\rm diss}$ (upper plot), and IR photon emission, $k_{\rm IR}$ (lower plot), as a function of the internal energy $E_{\rm int}$ and particle radius from 0.3 to 0.74\,nm, thinner to thicker lines, respectively. The dashed lines indicate particles with $N_{\rm C} = 23$ and 61 (see text for explanation).}
\label{fig_rates}
\end{figure}

\begin{figure} 
\centering
 \includegraphics[width=9.0cm]{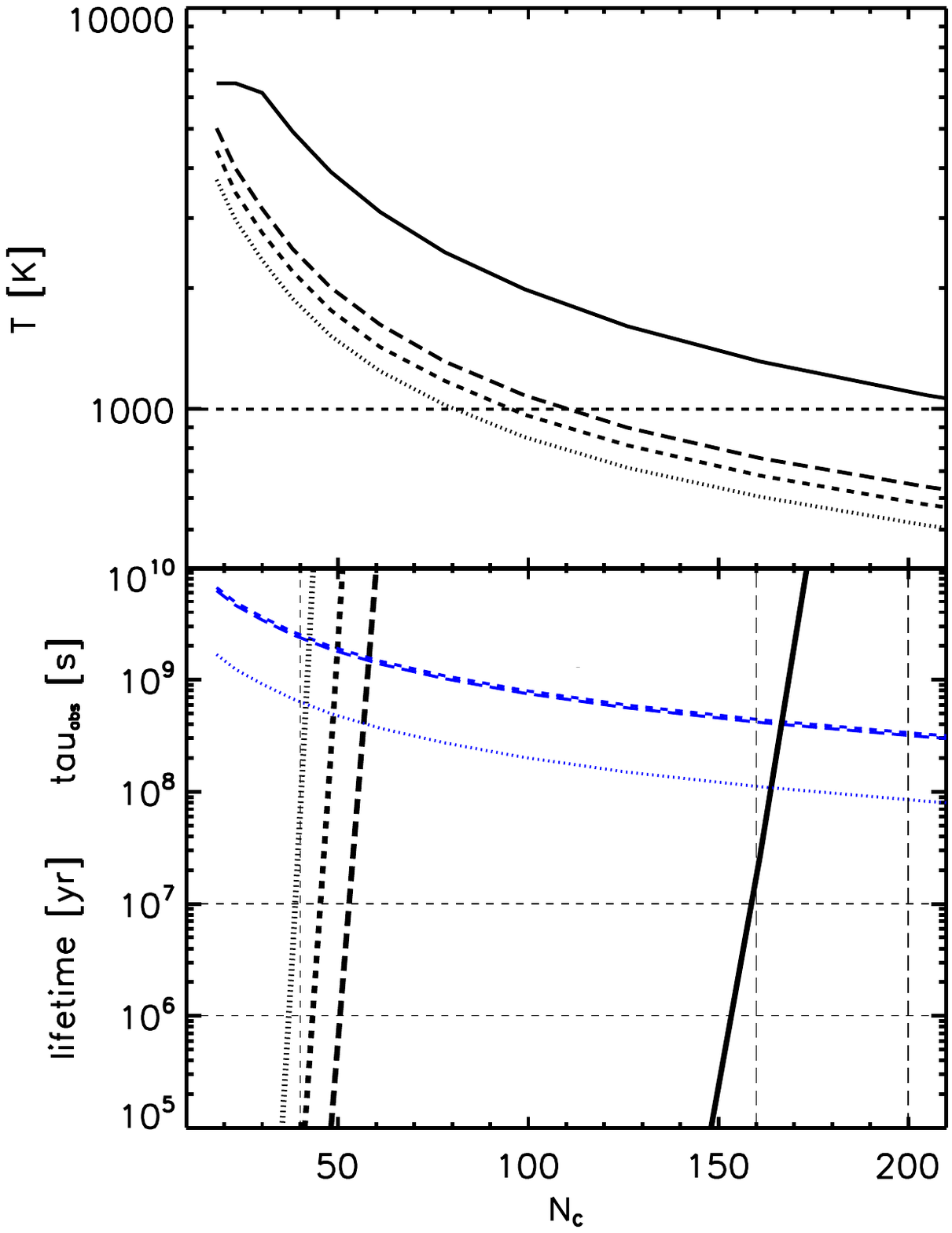}
 \caption{THEMIS a-C(:H) nanoparticle initial temperatures (upper plot), and in the lower plot photon absorption timescales (blue), and lifetimes (black) for the ISRF conditions 
 ($G_0$, $F_{\rm UV}$ $E_{h\nu}$, and $A_{\rm V}$) given in Table \ref{table_summary}.
 The thin vertical dashed lines indicate the THEMIS minimum grain size in the diffuse ISM ($a \sim 0.4-0.5$\,nm), and the derived limit for harsh PDR environments \cite[$a_- \lesssim 0.7-0.8$\,nm, IC63 and Orion Bar, ][]{2022A&A...666A..49S} and \cite[$a_- \sim 0.4$\,nm, Orion Bar, ][]{2024A&A...685A..76E}. The horizontal dashed lines mark the million year lifetime (lower plot) and $T=10^3$\,K  (upper plot).}
\label{fig_lifetimes}
\end{figure}

In our study we adopted $k_0 = 10^{16}$\,s$^{-1}$ \citep[similar to $k_0 = 1.4 \times 10^{16}$\,s$^{-1}$ of][]{2010A&A...510A..37M}\footnote{As these authors point out there are no experimental data for large PAHs or carbon clusters and so the exact values of $k_0$ are uncertain.} and $E_0 = 6.4$\,eV for a single bridging link.\footnote{We note that in their modelling \cite{2024A&A...692A.249B} derive a binding energy of $7.55 \pm 0.01$eV for C atoms within an amorphous carbonaceous structure but do not consider allyl adjacent bonding configurations.}
The ejection of an aromatic domain is less likely because its dissociation energy could be as high as $\sim 15$\,eV if for example, as shown previously, it were linked into the nanoparticle network by four aliphatic CC bonds. Hence, the motivation for considering that a-C nanoparticle destruction is instead triggered by the progressive and dissociative loss of the weaker bridging links.

\begin{table}[h]
\caption{Summary of parameters used in Fig. \ref{fig_lifetimes},  assuming $E_{\rm g} = 0.1$\,eV, $E_{\rm loss} = 6.4$\,eV, and $k_0 = 10^{16}$\,s$^{-1}$.  The various models are for: the diffuse ISM (DISM), cloud edges (edge) and an extreme PDR/H{\footnotesize II} region (HII).}
\begin{center}
\begin{tabular}{lccccc}
      &         &      &      &    &   \\[-0.35cm]
\hline
\hline
      &         &     &   &   &  \\[-0.35cm]
 model   &    $G_0$   &   $F_{\rm UV}$                &  $E_{h\nu}$   &  $A_{\rm V}$   &  line  \\[0.05cm]
              &                  &    [photons/cm$^{2}$/s]   &         [eV]       &  [mag]             &   style \\[0.05cm]
\hline
               &                    &              &       &     \\[-0.35cm]
DISM      &      1          &   $3 \times 10^7$      &     8     &    0.1    &  dotted       \\  
 edge     &      1          &   $1.2 \times 10^7$   &    10    &    0.5    &  short dash \\  
 edge     &      1          &   $7 \times 10^6$      &    12    &    0.5    &  long dash  \\  
HII      &  $10^4$    &   $7 \times 10^{10}$      &    30    &     0.1   &  solid          \\  
\hline
\hline
           &         &       &      &     &   \\[-0.25cm]
\end{tabular}
\end{center}
\label{table_summary}
\end{table}

The rate constants for dissociation and IR photon emission are shown in Fig. \ref{fig_rates} for grain radii from 0.3 to 0.74\,nm and can be compared with those of \cite{2013A&A...552A..15M}. In Fig. \ref{fig_rates} the dashed lines indicate the dissociation rate constants for particles with $N_{\rm C} = 23$ and 61, that is those closest in number of carbon atoms to the PAHs coronene, C$_{24}$H$_{12}$, and circumcorene, C$_{66}$H$_{20}$, presented in Fig. 3 of  \cite{2013A&A...552A..15M}. The dissociation rate constants calculated with the method presented here appear to be good agreement for the larger species but are significantly higher for smaller species. In other words we predict significantly higher photo-dissociation rates for the THEMIS a-C(:H) nanoparticles than for perfect planar PAHs, which is perhaps not surprising given the seemingly more fragile nature of the THEMIS nanograin structures (see Fig. \ref{fig_np_unfolded}). 

The nanoparticle UV photon absorption timescale is 
\begin{equation}
\tau_{\rm abs} = \left[ Q_{\rm abs}(E_{\rm g},a, \lambda) \ \pi a^2 \ F_{\rm UV} \ {\rm exp}(- 2.5 \, A_{\rm V}) \ G_0 \right]^{-1}
\end{equation}
where $Q_{\rm abs}(E_{\rm g},a, \lambda)$ is the absorption efficiency factor, $a$ is the nanoparticle radius, $F_{\rm UV}$ is the UV photon flux, the exponential factor is the UV extinction factor,  taken to be $\sim 2.5 \times A_{\rm V}$ \citep[assuming the average Galactic extinction curve  from][]{2007ApJ...663..320F}, and $G_0$ is the interstellar radiation field (ISRF) scaling factor. We take the nanoparticle photo-fragmentation lifetime, $t_{\rm pf}$, to be 
\begin{equation}
t_{\rm pf} = \frac{ 2 \, N_{\rm ar} \ \tau_{\rm abs} }{ P(n_{\rm max})  }
\end{equation}
where $N_{\rm ar}$ is the number of aromatic domains per nanoparticle the factor 2 is due to the equivalent need to liberate two aliphatic-olefinic bridges ($2 \times E_0 = 6.4$eV) or, that is, to break four  C$-$C bonds ($4 \times 3.2$eV) and thus to liberate an aromatic domain (see Fig. \ref{fig_np_unfolded} and Section \ref{sect_aromatics_network}).  This lifetime calculation makes the simplifying assumption that the particle properties are independent of photon absorption events, that is, the progressive erosion is not taken into account. With this assumption it is evident that the derived lifetimes are therefore upper limits. Nevertheless, this limitation is probably counteracted by processes that lead to particle annealing by bond (re-)formation and/or the accretion of atoms from the gas. It should, however, be pointed out that the dissociative erosion of nanoparticles leads to smaller particles that have even shorter lifetimes, as is clearly evident in Fig. \ref{fig_lifetimes}, because $t_{\rm pf}$ decreases steeply with $N_{\rm C}$.  

Some example results are shown in Fig. \ref{fig_lifetimes} for fluxes corresponding to $E_{h\nu} = 8$, 10 and 12eV photons, the dotted, short, and long dashed lines, respectively, with the local UV fluxes taken from \cite{2002ApJ...570..697H}. This figure shows that a million year lifetime is compatible with the minimum grain size, $a_- \sim 0.4$nm ($N_{\rm C} \simeq 40$), as per the THEMIS diffuse ISM model \citep{2013A&A...558A..62J,2017A&A...602A..46J}. Also shown is an illustrative example that leads to the photo-destruction of all particles with less than $160-200$ C atoms \cite[$a \lesssim 0.7-0.8$nm, as per][]{2022A&A...666A..49S}, which requires an excitation or absorbed energy per particle of  $\simeq 30$eV. Such a high excitation energy could be achieved by hard photons ($E_{h\nu} = 30$eV) in a harsh ionising ISRF or through the effects of multi-photon absorption events \cite[e.g.][]{2013A&A...552A..15M}, indicating that the destruction of a-C nanoparticles with hundreds of carbon atoms requires extremely high internal excitation energies. In their detailed JWST spectroscopic data analysis \cite{2024A&A...685A..76E} determine a minimum grain size ($a_- \sim 0.4$nm) for the Orion Bar that is compatible with the THEMIS diffuse ISM model \citep{2013A&A...558A..62J,2017A&A...602A..46J} but significantly  smaller that that found by \cite{2022A&A...666A..49S} using {\em Spitzer} and {\em Herschel} photometric data. 

We find that for our adopted photo-fragmentation scheme it is both the intensity (possibly leading to multi-photon absorption events) and the hardness (the absorption of extreme UV photons) of the ISRF that are critical. In these calculations the nanoparticle lifetimes are determined by assuming the breaking of four C$-$C aliphatic-olefinic bridge bonds per aromatic domain equivalent to the ejection of two complete aliphatic-olefinic bridges. 

Unfortunately the values of $k_0$ and $E_0$ that are adopted here are rather uncertain for a-C(:H) nanoparticles, and more so than for PAHs, for which there is some experimental evidence. Hence the current results should be regarded as order of magnitude estimates for the photo-fragmentation rates of the THEMIS a-C(:H) nanoparticles until such time as relevant experimental data and/or more rigorous models become available. \\

\section{Coulomb fragmentation}
\label{sect_DCF}

In amorphous semi-conductors such as a-C(:H), conduction is excitation-induced with the charge carriers (electrons or holes) bound to, and promoted between, the aromatic domains. In finite-sized a-C(:H) particles the charges will therefore be localised on the aromatic sub-structures and, in multiply-charged a-C(:H) nanoparticles with a total charge $\geqslant 2e$, the charge carriers will tend to reside on different aromatic domains because of Coulomb repulsion. The maximum distance between the charged aromatic domains, in a cage-like a-C(:H) nanoparticle, is the diameter of the particle, which is likely to be of the order of $0.5-1$\,nm for the types of particles where Coulomb fragmentation might be important \citep[e.g.][]{2012ApJ...761...35M}. 

From the above discussion and calculations we can reasonably assume (on a statistical basis) that a typical aromatic domain consists of $\simeq 14$ C atoms ($\equiv N_{\rm R} = 3$) and that it is linked within the a-C(:H) grain network by four bridging structures, which will dissociate upon Multi-cation induced Coulomb Fragmentation (MCF) and such an occurence will result in the loss of $\simeq 20$ C atoms from the particle. In this case the extra-aromatic C atoms result from the disruption of some of the aliphatic-olefinic inter-domain bridges \cite[e.g.][]{2015A&A...581A..92J}. 

As shown earlier, the number of aromatic domains in these shell-like structures increases with radius (see Eq. \ref{eq_ar_separation}). Nevertheless, for a given size particle, as the number of charges increases the distance between the charge carrying domains must necessarily decrease. In fact the minimum distance between the aromatic domains in an a-C(:H) nanoparticle is $\simeq 0.4-0.5$nm.  

Within an a-C(:H) nanoparticle the Coulomb force, $F_{\rm C}$, and electrostatic potential, $U_{\rm C}$, between two charged aromatic, sub-particle domains, with charges $Z_1$ and $Z_2$, are given by
\begin{equation}
F_{\rm C} = \frac{ Z_1 \ Z_2 \ e^2 }{ d_{\rm ar}^2} \ \ \ \ \ {\rm and} \ \ \ \ \ U_{\rm C} = \frac{ Z_1 \ Z_2 \ e^2 }{ d_{\rm ar} },
\end{equation}
respectively, where $e$ can be expressed in e.s.u. and $d_{\rm ar}$ in cm. MCF can occur if the Coulomb repulsive force between these charged aromatic domains exceeds the tensile strength that holds either of them within the network, that is, if the repulsive force is greater than the sum of the bond energies binding one of the domains within the contiguous network. As shown above, assuming an aromatic domain is bound by four single aliphatic bridge bonds (see Fig. \ref{fig_np_unfolded} and Section \ref{sect_sized_aromatics_network}), the binding energy for an aromatic domain is then $\simeq 4 \, E_{\rm br}$, where $E_{\rm br}\simeq 3.2$\,eV is the binding energy of a single aliphatic bond that links an aromatic domains into the network. Thus, the energy required to liberate an aromatic domain is of the order of 12.8\,eV. 

If we assume that particles with radii of 0.2, 0.25, 0.3, 0.35, and 0.4\,nm contain, on average, 2, 2, 3, 3, and 4 aromatic domains, and that the distance between these domains is $\simeq 0.4$\,nm, then charges of +2 or +3 on adjacent aromatic domains are enough to drive one of those domains from the nanoparticle and therefore to significantly disrupt it (i.e. $E > 12.8$\,eV). To arrive at this catastrophic situation requires that the other aromatic domains in the particle are almost equally charged, leading to a situation in which the total charge state for MCF increases with particle size. For the above parameters, MCF would occur for particles with radii of 0.2 and 0.25\,nm with each of the two constituent aromatic domains charged to +2 or +3 or a total particle charge of +4 or +6, respectively. Where the nanoparticles contains three [four] aromatic domains the charge states required for MCF would be +2/+3/+3 (total +8) [+2/+2/+3/+3 (total +10)].  
These example charge states would appear to be significantly higher than those typically found on interstellar grains  \cite[e.g.][]{2019MNRAS.485.1220I},  even in PDRs, which would seem to argue against charge-induced effects being at the heart of carbonaceous nanoparticle destruction in high excitation regions. 

Given that the nanoparticle charge states required to induce MCF are rather extreme, it would appear that destructive charge effects on nanoparticles are subordinate to the photo-fragmentation effects discussed in the preceding sections.

\section{Conclusions}
\label{sect_conclusions}

The heterogeneous structural properties of semiconducting, hydrogenated amorphous carbon materials, a-C(:H), pose a challenge to the investigation of their survival in the ISM and in PDRs, especially in the most highly excitation regions. Following a detailed study of the chemical bonding makeup of a-C(:H) nanoparticles, that takes their intrinsic heterogeneity into account, this work models their photo- and charge-stability in the ISM by developing a detailed understanding of the nature of their constituent aliphatic-olefinic and aromatic domain sub-structures. The important developmental aspects and major conclusions of this work are:

\begin{enumerate}

\item
Using a statistical approach we estimated the aromatic domain sizes, their size distribution, how they are bonded into a contiguous 3D a-C(:H) network, and where they are found within these structures as a function of the particle size. 

\item
We find that, as constrained by network structure and H atom fraction considerations, single aromatic ring systems ($N_{\rm R} = 1$ with $N_{\rm C} = 6$, i.e. benzene-like rings) are not favoured in the smallest nanoparticles ($N_{\rm C} < 80$). 
In general, aromatic domains with $N_{\rm R} = 2-3$ are the most common in a-C(:H) nanoparticle structures containing $40-80$ carbon atoms. 

\item
About one half of the peripheral bonding sites on aromatic domains are network bonded, and about one half are hydrogenated, exhibiting aromatic CH bonds, which will, statistically, tend to solo-type CH configurations. 

\item
Aromatic domains are linked into the 3D structure via aliphatic-olefinic bridges or chains with typically $4-6$ carbon atoms. Non-aromatic cyclic domain linking structures are also possible and would appear to be essential in explaining the aromatisation of hydrogenated amorphous carbons. 

\item
In bulk a-C(:H) materials, where a power law aromatic domain size distribution favours the smaller domains and the optical properties are determined by the largest. In nanoparticles the aromatic domains tend towards the largest possible. 

\item
In a-C(:H) nanoparticles the sizes (diameters) of the aromatic domains must be, at most, similar to the particle radius in order to preserve a contiguous 3D structure. This makes sense physically because sub-nm sized a-C(:H) particles tend to exhibit cage- or shell-like structures.\footnote{This can perhaps best be illustrated by imagining such a shell as being approximately hexagonal or octagonal in cross-section. In this case the linear dimension of the peripheral domains, that is the edges, are similar to the shell radius.}

\item
Interestingly, and perhaps counter intuitively, within mixed aliphatic, olefinic, and aromatic 3D structural networks, the weakest bonds are not the CH bonds ($D_{\rm CH} \sim 4.3$eV) nor the aliphatic CC bonds ($D_{\rm CC} \sim 3.8$eV). This prize goes to allyl-adjacent aliphatic C-$\wr$-C bonds (--C-$\wr$-C--C=C--, $D_{\rm C-\wr-C} \sim 3.2$eV) in mixed aliphatic and olefinic chains. 

\item
Aliphatic chains (e.g. --C--C--C--C--) between aromatic domains are more resistant to photodissociation than their partially dehydrogenated products (e.g. --C--C--C=C--). Dehydrogenation therefore tends to weaken a-C(:H) nanoparticle structures before strengthening them by aromatisation.  

\item
We considered the effects of thermal excitation and photo-dissociation on a-C(:H) nanoparticles using an approach based on photo-thermo-dissociation methodologies, allowing us to predict the fragmentation timescales for the UV photon-driven photodissociation of CC bonds. 

\item
We estimated a-C(:H) nanoparticle lifetimes in the ISM and in PDRs of the order of $10^6-10^7$yr, depending on the local ISRF, for particles with radii of $0.4-0.5$nm ($N_{\rm C} = 40-60$), sizes that are consistent with the smallest nanoparticles predicted by THEMIS. 

\item
As concluded in studies of PAH survival, we found that particles with less than $\lesssim 50$ carbon atoms are unstable in the diffuse ISM where the mean UV photn energy is $8-10$eV. 

\item
In H{\footnotesize II} regions, where hard UV photons ($E_{\rm h \nu} > 13.6$eV) are present and multiple UV photon absorption events can occur, only particles with radii greater than 0.7nm ($\gtrsim 150$ carbon atoms) are likely to survive.

\item
Energetic photon absorption leading to charge-induced Coulomb fragmentation was shown to be an unimportant dust destruction process, except in the case of extremely large grain charges on small grains where each of the constituent aromatic domains carry charges of $\geqslant +2$ or +3.

\end{enumerate}


\begin{acknowledgements} 
The authors would like to thank the referee for a very thorough, careful and detailed review, which helped to greatly improve the paper. 
\end{acknowledgements}



\bibliographystyle{aa} 
\bibliography{Ant_bibliography.bib} 

@article{2024A&A...692A.249B,
	adsurl = {https://ui.adsabs.harvard.edu/abs/2024A&A...692A.249B},
	archiveprefix = {arXiv},
	author = {{Bossion}, D. and {Sarangi}, A. and {Aalto}, S. and {Esmerian}, C. and {Hashemi}, S.~R. and {Knudsen}, K.~K. and {Vlemmings}, W. and {Nyman}, G.},
	doi = {10.1051/0004-6361/202452362},
	eid = {A249},
	eprint = {2411.06125},
	journal = {\aap},
	month = dec,
	pages = {A249},
	primaryclass = {astro-ph.GA},
	title = {{Accurate sticking coefficient calculation for carbonaceous dust growth through accretion and desorption in astrophysical environments}},
	volume = {692},
	year = 2024}

@article{2019MNRAS.485.1220I,
	adsurl = {https://ui.adsabs.harvard.edu/abs/2019MNRAS.485.1220I},
	archiveprefix = {arXiv},
	author = {{Ib{\'a}{\~n}ez-Mej{\'\i}a}, Juan C. and {Walch}, Stefanie and {Ivlev}, Alexei V. and {Clarke}, Seamus and {Caselli}, Paola and {Joshi}, Prabesh R.},
	doi = {10.1093/mnras/stz207},
	eprint = {1812.08281},
	journal = {\mnras},
	month = may,
	number = {1},
	pages = {1220-1247},
	primaryclass = {astro-ph.GA},
	title = {{Dust charge distribution in the interstellar medium}},
	volume = {485},
	year = 2019}

@article{2024A&A...684A..34Y,
	adsurl = {https://ui.adsabs.harvard.edu/abs/2024A&A...684A..34Y},
	archiveprefix = {arXiv},
	author = {{Ysard}, N. and {Jones}, A.~P. and {Guillet}, V. and {Demyk}, K. and {Decleir}, M. and {Verstraete}, L. and {Choubani}, I. and {Miville-Desch{\^e}nes}, M. -A. and {Fanciullo}, L.},
	doi = {10.1051/0004-6361/202348391},
	eid = {A34},
	eprint = {2401.07739},
	journal = {\aap},
	month = apr,
	pages = {A34},
	primaryclass = {astro-ph.GA},
	title = {{THEMIS 2.0: A self-consistent model for dust extinction, emission, and polarisation}},
	volume = {684},
	year = 2024}

@article{2024A&A...685A..76E,
	adsurl = {https://ui.adsabs.harvard.edu/abs/2024A&A...685A..76E},
	archiveprefix = {arXiv},
	author = {{Elyajouri}, M. and {Ysard}, N. and {Abergel}, A. and {Habart}, E. and {Verstraete}, L. and {Jones}, A. and {Juvela}, M. and {Schirmer}, T. and {Meshaka}, R. and {Dartois}, E. and {Lebourlot}, J. and {Rouill{\'e}}, G. and {Onaka}, T. and {Peeters}, E. and {Bern{\'e}}, O. and {Alarc{\'o}n}, F. and {Bernard-Salas}, J. and {Buragohain}, M. and {Cami}, J. and {Canin}, A. and {Chown}, R. and {Demyk}, K. and {Gordon}, K. and {Kannavou}, O. and {Kirsanova}, M. and {Madden}, S. and {Paladini}, R. and {Pendleton}, Y. and {Salama}, F. and {Schroetter}, I. and {Sidhu}, A. and {R{\"o}llig}, M. and {Trahin}, B. and {Van De Putte}, D.},
	doi = {10.1051/0004-6361/202348728},
	eid = {A76},
	eprint = {2401.01221},
	journal = {\aap},
	month = may,
	pages = {A76},
	primaryclass = {astro-ph.GA},
	title = {{PDRs4All. V. Modelling the dust evolution across the illuminated edge of the Orion Bar}},
	volume = {685},
	year = 2024}

@article{2018ARA&A..56..673G,
	adsurl = {https://ui.adsabs.harvard.edu/abs/2018ARA&A..56..673G},
	archiveprefix = {arXiv},
	author = {{Galliano}, Fr{\'e}d{\'e}ric and {Galametz}, Maud and {Jones}, Anthony P.},
	doi = {10.1146/annurev-astro-081817-051900},
	eprint = {1711.07434},
	journal = {\araa},
	month = sep,
	pages = {673-713},
	primaryclass = {astro-ph.GA},
	title = {{The Interstellar Dust Properties of Nearby Galaxies}},
	volume = {56},
	year = 2018}

@article{2022A&A...666A..49S,
	adsurl = {https://ui.adsabs.harvard.edu/abs/2022A&A...666A..49S},
	archiveprefix = {arXiv},
	author = {{Schirmer}, T. and {Ysard}, N. and {Habart}, E. and {Jones}, A.~P. and {Abergel}, A. and {Verstraete}, L.},
	doi = {10.1051/0004-6361/202243635},
	eid = {A49},
	eprint = {2209.04282},
	journal = {\aap},
	month = oct,
	pages = {A49},
	primaryclass = {astro-ph.GA},
	title = {{Nano-grain depletion in photon-dominated regions}},
	volume = {666},
	year = 2022}

@article{2016MNRAS.459.2751K,
	adsurl = {https://ui.adsabs.harvard.edu/abs/2016MNRAS.459.2751K},
	archiveprefix = {arXiv},
	author = {{Kimura}, Hiroshi},
	doi = {10.1093/mnras/stw820},
	eprint = {1604.03664},
	journal = {\mnras},
	month = jul,
	number = {3},
	pages = {2751-2761},
	primaryclass = {astro-ph.EP},
	title = {{On the photoelectric quantum yield of small dust particles}},
	volume = {459},
	year = 2016}

@article{2013A&A...552A..15M,
	adsurl = {https://ui.adsabs.harvard.edu/abs/2013A&A...552A..15M},
	archiveprefix = {arXiv},
	author = {{Montillaud}, J. and {Joblin}, C. and {Toublanc}, D.},
	doi = {10.1051/0004-6361/201220757},
	eid = {A15},
	eprint = {1301.6507},
	journal = {\aap},
	month = apr,
	pages = {A15},
	primaryclass = {astro-ph.GA},
	title = {{Evolution of polycyclic aromatic hydrocarbons in photodissociation regions. Hydrogenation and charge states}},
	volume = {552},
	year = 2013}

@article{2017CTC...1115...45,
	author = {{Quan-De Wang}},
	journal = {Computational and Theoretical Chemistry},
	pages = {45-55},
	title = {{Theoretical studies of unimolecular thermal decomposition reactions of n-hexane and n-hexene isomers}},
	volume = {1115},
	year = {2017}}

@article{1999ApJ...512..500J,
	adsurl = {https://ui.adsabs.harvard.edu/abs/1999ApJ...512..500J},
	author = {{Jochims}, H.~W. and {Baumg{\"a}rtel}, H. and {Leach}, S.},
	doi = {10.1086/306752},
	journal = {\apj},
	month = feb,
	number = {1},
	pages = {500-510},
	title = {{Structure-dependent Photostability of Polycyclic Aromatic Hydrocarbon Cations: Laboratory Studies and Astrophysical Implications}},
	volume = {512},
	year = 1999}

@article{2021A&A...649A.148S,
	adsurl = {https://ui.adsabs.harvard.edu/abs/2021A&A...649A.148S},
	archiveprefix = {arXiv},
	author = {{Schirmer}, T. and {Habart}, E. and {Ysard}, N. and {Bron}, E. and {Le Bourlot}, J. and {Verstraete}, L. and {Abergel}, A. and {Jones}, A.~P. and {Roueff}, E. and {Le Petit}, F.},
	doi = {10.1051/0004-6361/202140671},
	eid = {A148},
	eprint = {2104.08204},
	journal = {\aap},
	month = may,
	pages = {A148},
	primaryclass = {astro-ph.GA},
	title = {{Influence of the nano-grain depletion in photon-dominated regions. Application to the gas physics and chemistry in the Horsehead}},
	volume = {649},
	year = 2021}

@article{2021A&A...649A..84H,
	adsurl = {https://ui.adsabs.harvard.edu/abs/2021A&A...649A..84H},
	archiveprefix = {arXiv},
	author = {{Habart}, E. and {Bout{\'e}raon}, T. and {Brauer}, R. and {Ysard}, N. and {Pantin}, E. and {Marchal}, A. and {Jones}, A.~P.},
	doi = {10.1051/0004-6361/201936388},
	eid = {A84},
	eprint = {2011.13773},
	journal = {\aap},
	month = may,
	pages = {A84},
	primaryclass = {astro-ph.GA},
	title = {{Spatial distribution of the aromatic and aliphatic carbonaceous nanograin features in the protoplanetary disk around HD 100546}},
	volume = {649},
	year = 2021}

@article{2020A&A...639A.144S,
	adsurl = {https://ui.adsabs.harvard.edu/abs/2020A&A...639A.144S},
	archiveprefix = {arXiv},
	author = {{Schirmer}, T. and {Abergel}, A. and {Verstraete}, L. and {Ysard}, N. and {Juvela}, M. and {Jones}, A.~P. and {Habart}, E.},
	doi = {10.1051/0004-6361/202037937},
	eid = {A144},
	eprint = {2003.05902},
	journal = {\aap},
	month = jul,
	pages = {A144},
	primaryclass = {astro-ph.GA},
	title = {{Dust evolution across the Horsehead nebula}},
	volume = {639},
	year = 2020}

@article{2019A&A...627A..38J,
	adsurl = {https://ui.adsabs.harvard.edu/abs/2019A&A...627A..38J},
	archiveprefix = {arXiv},
	author = {{Jones}, A.~P. and {Ysard}, N.},
	doi = {10.1051/0004-6361/201935532},
	eid = {A38},
	eprint = {1906.01382},
	journal = {\aap},
	month = {Jul},
	pages = {A38},
	primaryclass = {astro-ph.GA},
	title = {{The essential elements of dust evolution. A viable solution to the interstellar oxygen depletion problem?}},
	volume = {627},
	year = {2019}}

@article{2019A&A...569A.100B,
	author = {Bout{\'e}raon, T and Habart, E. and Ysard, N and Jones, A P and Dartois, E and Pino, T},
	journal = {\aap},
	pages = {A135},
	title = {{Nano carbon dust emission in proto-planetary disks: the aliphatic-aromatic components}},
	volume = {623},
	year = {2019}}

@article{wang_etal_2001,
	author = {B.-C. Wang and H.-W. Wang and J.-C. Chang and H.-C. Tso and Y.-M. Chou},
	journal = {Journal of Molecular Structure: THEOCHEM},
	pages = {171},
	title = {More spherical large fullerenes and multi-layer fullerene cages},
	volume = {540},
	year = {2001}}

@article{2008ApJ...672..214G,
	adsurl = {http://cdsads.u-strasbg.fr/abs/2008ApJ...672..214G},
	archiveprefix = {arXiv},
	author = {{Galliano}, F. and {Dwek}, E. and {Chanial}, P.},
	doi = {10.1086/523621},
	eid = {214-243},
	eprint = {0708.0790},
	journal = {\apj},
	month = jan,
	pages = {214-243},
	title = {{Stellar Evolutionary Effects on the Abundances of Polycyclic Aromatic Hydrocarbons and Supernova-Condensed Dust in Galaxies}},
	volume = 672,
	year = 2008}

@article{2017A&A...602A..46J,
	adsurl = {http://cdsads.u-strasbg.fr/abs/2017A%26A...602A..46J},
	author = {{Jones}, A.~P. and {K{\"o}hler}, M. and {Ysard}, N. and {Bocchio}, M. and {Verstraete}, L.},
	doi = {10.1051/0004-6361/201630225},
	eid = {A46},
	journal = {\aap},
	month = jun,
	pages = {A46},
	title = {{The global dust modelling framework THEMIS}},
	volume = 602,
	year = 2017}

@article{2016RSOS....360224J,
	adsurl = {http://cdsads.u-strasbg.fr/abs/2016RSOS....360224J},
	author = {{Jones}, A.~P.},
	doi = {10.1098/rsos.160224},
	eid = {160224},
	journal = {Royal Society Open Science},
	month = dec,
	pages = {160224},
	title = {{Dust evolution, a global view: III. Core/mantle grains, organic nano-globules, comets and surface chemistry}},
	volume = 3,
	year = 2016}

@article{2016RSOS....360223J,
	adsurl = {http://cdsads.u-strasbg.fr/abs/2016RSOS....360223J},
	author = {{Jones}, A.~P.},
	doi = {10.1098/rsos.160223},
	eid = {160223},
	journal = {Royal Society Open Science},
	month = dec,
	pages = {160223},
	title = {{Dust evolution, a global view: II. Top-down branching, nanoparticle fragmentation and the mystery of the diffuse interstellar band carriers}},
	volume = 3,
	year = 2016}

@article{2016RSOS....360221J,
	adsurl = {http://cdsads.u-strasbg.fr/abs/2016RSOS....360221J},
	author = {{Jones}, A.~P.},
	doi = {10.1098/rsos.160221},
	eid = {160221},
	journal = {Royal Society Open Science},
	month = dec,
	pages = {160221},
	title = {{Dust evolution, a global view I. Nanoparticles, nascence, nitrogen and natural selection {\ldots} joining the dots}},
	volume = 3,
	year = 2016}

@article{2012A&A...548A..40C,
	author = {Carpentier, Y and F{\'e}raud, G and Dartois, E and Brunetto, R and Charon, E and Cao, A-T and d'Hendecourt, L and Br{\'e}chignac, Ph and Rouzaud, J N and Pino, T},
	journal = {\aap},
	month = dec,
	pages = {A40},
	title = {{Nanostructuration of carbonaceous dust as seen through the positions of the 6.2 and 7.7 $\mu$m AIBs}},
	volume = {548},
	year = {2012}}

@article{2006ApJ...645.1188W,
	adsurl = {http://cdsads.u-strasbg.fr/abs/2006ApJ...645.1188W},
	author = {{Weingartner}, J.~C. and {Draine}, B.~T. and {Barr}, D.~K.},
	doi = {10.1086/504420},
	eprint = {astro-ph/0601296},
	journal = {\apj},
	month = jul,
	pages = {1188-1197},
	title = {{Photoelectric Emission from Dust Grains Exposed to Extreme Ultraviolet and X-Ray Radiation}},
	volume = 645,
	year = 2006}

@article{1989A&A...213..351L,
	adsurl = {http://cdsads.u-strasbg.fr/abs/1989A%26A...213..351L},
	author = {{Leger}, A. and {D'Hendecourt}, L. and {Boissel}, P. and {Desert}, F.~X.},
	journal = {\aap},
	month = apr,
	pages = {351-359},
	title = {{Photo-thermo-dissociation. I - A general mechanism for destroying molecules}},
	volume = 213,
	year = 1989}

@article{2016A&A...588A..43J,
	adsurl = {http://adsabs.harvard.edu/abs/2016A%26A...588A..43J},
	archiveprefix = {arXiv},
	author = {{Jones}, A.~P. and {K{\"o}hler}, M. and {Ysard}, N. and {Dartois}, E. and {Godard}, M. and {Gavilan}, L.},
	doi = {10.1051/0004-6361/201527488},
	eid = {A43},
	eprint = {1602.00538},
	journal = {\aap},
	month = apr,
	pages = {A43},
	title = {{Mantle formation, coagulation, and the origin of cloud/core shine. I. Modelling dust scattering and absorption in the infrared}},
	volume = 588,
	year = 2016}

@article{2009SSRv..143..311S,
	adsurl = {http://cdsads.u-strasbg.fr/abs/2009SSRv..143..311S},
	archiveprefix = {arXiv},
	author = {{Slavin}, J.~D.},
	doi = {10.1007/s11214-008-9342-3},
	eprint = {0804.0161},
	journal = {\ssr},
	month = mar,
	pages = {311-322},
	title = {{The Origins and Physical Properties of the Complex of Local Interstellar Clouds}},
	volume = 143,
	year = 2009}

@article{2015A&A...584A.123A,
	adsurl = {http://cdsads.u-strasbg.fr/abs/2015A%26A...584A.123A},
	author = {{Alata}, I. and {Jallat}, A. and {Gavilan}, L. and {Chabot}, M. and {Cruz-Diaz}, G.~A. and {Munoz Caro}, G.~M. and {B{\'e}roff}, K. and {Dartois}, E.},
	doi = {10.1051/0004-6361/201526368},
	eid = {A123},
	journal = {\aap},
	month = dec,
	pages = {A123},
	title = {{Vacuum ultraviolet of hydrogenated amorphous carbons. II. Small hydrocarbons production in Photon Dominated Regions}},
	volume = 584,
	year = 2015}

@article{2015A&A...577A.110Y,
	adsurl = {http://cdsads.u-strasbg.fr/abs/2015A%26A...577A.110Y},
	archiveprefix = {arXiv},
	author = {{Ysard}, N. and {K{\"o}hler}, M. and {Jones}, A. and {Miville-Desch{\^e}nes}, M.-A. and {Abergel}, A. and {Fanciullo}, L.},
	doi = {10.1051/0004-6361/201425523},
	eid = {A110},
	eprint = {1503.07435},
	journal = {\aap},
	month = may,
	pages = {A110 (593, C4)},
	title = {{Dust variations in the diffuse interstellar medium: constraints on Milky Way dust from Planck-HFI observations}},
	volume = 577,
	year = 2015}

@article{1995Ap&SS.224..417A,
	adsurl = {http://cdsads.u-strasbg.fr/abs/1995Ap%26SS.224..417A},
	author = {{Allain}, T. and {Sedlmayr}, E. and {Leach}, S.},
	doi = {10.1007/BF00667884},
	journal = {\apss},
	month = feb,
	pages = {417-418},
	title = {{Formation and Photodestruction of PAHs}},
	volume = 224,
	year = 1995}

@article{1996A&A...305..602A,
	adsurl = {http://cdsads.u-strasbg.fr/abs/1996A%26A...305..602A},
	author = {{Allain}, T. and {Leach}, S. and {Sedlmayr}, E.},
	journal = {\aap},
	month = jan,
	pages = {602},
	title = {{Photodestruction of PAHs in the interstellar medium. I. Photodissociation rates for the loss of an acetylenic group.}},
	volume = 305,
	year = 1996}

@article{1996A&A...305..616A,
	adsurl = {http://cdsads.u-strasbg.fr/abs/1996A%26A...305..616A},
	author = {{Allain}, T. and {Leach}, S. and {Sedlmayr}, E.},
	journal = {\aap},
	month = jan,
	pages = {616},
	title = {{Photodestruction of PAHs in the interstellar medium. II. Influence of the states of ionization and hydrogenation.}},
	volume = 305,
	year = 1996}

@article{2011A&A...526A..52M,
	author = {{Micelotta}, E.~R. and {Jones}, A.~P. and {Tielens}, A.~G.~G.~M.},
	journal = {\aap},
	pages = {A52},
	title = {{Polycyclic aromatic hydrocarbon processing by cosmic rays}},
	volume = {526},
	year = {2011}}

@article{2014A&A...570A..32B,
	adsurl = {http://cdsads.u-strasbg.fr/abs/2014A%26A...570A..32B},
	author = {{Bocchio}, M. and {Jones}, A.~P. and {Slavin}, J.~D.},
	doi = {10.1051/0004-6361/201424368},
	eid = {A32},
	journal = {\aap},
	month = oct,
	pages = {A32},
	title = {{A re-evaluation of dust processing in supernova shock waves}},
	volume = 570,
	year = 2014}

@article{2015A&A...581A..92J,
	adsurl = {http://cdsads.u-strasbg.fr/abs/2015A%26A...581A..92J},
	archiveprefix = {arXiv},
	author = {{Jones}, A.~P. and {Habart}, E.},
	doi = {10.1051/0004-6361/201526487},
	eid = {A92},
	eprint = {1507.08534},
	journal = {\aap},
	month = sep,
	pages = {A92},
	title = {{H$_{2}$ formation via the UV photo-processing of a-C:H nano-particles}},
	volume = 581,
	year = 2015}

@article{2012A&A...542A..98J,
	adsurl = {http://adsabs.harvard.edu/abs/2012A%26A...542A..98J},
	author = {{Jones}, A.~P.},
	doi = {10.1051/0004-6361/201118483},
	eid = {A98},
	journal = {\aap},
	month = jun,
	pages = {A98 (545, C3)},
	title = {{Variations on a theme - the evolution of hydrocarbon solids. III. Size-dependent properties - the optEC$_{(s)}$(a) model}},
	volume = 542,
	year = 2012}

@article{2012A&A...540A...2J,
	adsurl = {http://adsabs.harvard.edu/abs/2012A%26A...540A...2J},
	author = {{Jones}, A.~P.},
	doi = {10.1051/0004-6361/201117624},
	eid = {A2},
	journal = {\aap},
	month = apr,
	pages = {A2 (545, C2)},
	title = {{Variations on a theme - the evolution of hydrocarbon solids. II. Optical property modelling - the optEC$_{(s)}$ model}},
	volume = 540,
	year = 2012}

@article{2012A&A...540A...1J,
	adsurl = {http://adsabs.harvard.edu/abs/2012A%26A...540A...1J},
	author = {{Jones}, A.~P.},
	doi = {10.1051/0004-6361/201117623},
	eid = {A1},
	journal = {\aap},
	month = apr,
	pages = {A1},
	title = {{Variations on a theme - the evolution of hydrocarbon solids. I. Compositional and spectral modelling - the eRCN and DG models}},
	volume = 540,
	year = 2012}

@article{2015ApJ...800L..33G,
	adsurl = {http://cdsads.u-strasbg.fr/abs/2015ApJ...800L..33G},
	archiveprefix = {arXiv},
	author = {{Guzm{\'a}n}, V.~V. and {Pety}, J. and {Goicoechea}, J.~R. and {Gerin}, M. and {Roueff}, E. and {Gratier}, P. and {{\"O}berg}, K.~I.},
	doi = {10.1088/2041-8205/800/2/L33},
	eid = {L33},
	eprint = {1502.02325},
	journal = {\apjl},
	month = feb,
	pages = {L33},
	title = {{Spatially Resolved L-C$_{3}$H$^{+}$ Emission in the Horsehead Photodissociation Region: Further Evidence for a Top-Down Hydrocarbon Chemistry}},
	volume = 800,
	year = 2015}

@article{2014A&A...565L...9K,
	adsurl = {http://cdsads.u-strasbg.fr/abs/2014A%26A...565L...9K},
	archiveprefix = {arXiv},
	author = {{K{\"o}hler}, M. and {Jones}, A. and {Ysard}, N.},
	doi = {10.1051/0004-6361/201423985},
	eid = {L9},
	eprint = {1405.4208},
	journal = {\aap},
	month = may,
	pages = {L9},
	title = {{A hidden reservoir of Fe/FeS in interstellar silicates?}},
	volume = 565,
	year = 2014}

@article{Faraday_Disc_paper_2014,
	author = {{Jones}, A.~P. and {Ysard}, N. and {K\"ohler}, M. and {Fanciullo}, L. and {Bocchio}, M. and {Micelotta}, E.~R.and {Verstraete}, L. and {Guillet}, V.},
	journal = {RSC Faraday Discuss.},
	pages = {313},
	title = {{The cycling of carbon into and out of dust}},
	volume = {168},
	year = {2014}}

@article{2014A&A...569A.119A,
	adsurl = {http://cdsads.u-strasbg.fr/abs/2014A%26A...569A.119A},
	author = {{Alata}, I. and {Cruz-Diaz}, G.~A. and {Mu{\~n}oz Caro}, G.~M. and {Dartois}, E.},
	doi = {10.1051/0004-6361/201323118},
	eid = {A119},
	journal = {\aap},
	month = sep,
	pages = {A119},
	title = {{Vacuum ultraviolet photolysis of hydrogenated amorphous carbons . I. Interstellar H$_{2}$ and CH$_{4}$ formation rates}},
	volume = 569,
	year = 2014}

@article{2013A&A...558A..62J,
	adsurl = {http://cdsads.u-strasbg.fr/abs/2013A%26A...558A..62J},
	author = {{Jones}, A.~P. and {Fanciullo}, L. and {K{\"o}hler}, M. and {Verstraete}, L. and {Guillet}, V. and {Bocchio}, M. and {Ysard}, N.},
	doi = {10.1051/0004-6361/201321686},
	eid = {A62},
	journal = {\aap},
	month = oct,
	pages = {A62},
	title = {{The evolution of amorphous hydrocarbons in the ISM: dust modelling from a new vantage point}},
	volume = 558,
	year = 2013}

@article{2013A&A...555A..39J,
	adsurl = {http://cdsads.u-strasbg.fr/abs/2013A%26A...555A..39J},
	author = {{Jones}, A.~P.},
	doi = {10.1051/0004-6361/201321687},
	eid = {A39},
	journal = {\aap},
	month = jul,
	pages = {A39},
	title = {{Heteroatom-doped hydrogenated amorphous carbons, a-C:H:X. ''Volatile'' silicon, sulphur and nitrogen depletion, blue photoluminescence, diffuse interstellar bands and ferro-magnetic carbon grain connections}},
	volume = 555,
	year = 2013}

@article{2012A&A...548A..68P,
	adsurl = {http://cdsads.u-strasbg.fr/abs/2012A%26A...548A..68P},
	archiveprefix = {arXiv},
	author = {{Pety}, J. and {Gratier}, P. and {Guzm{\'a}n}, V. and {Roueff}, E. and {Gerin}, M. and {Goicoechea}, J.~R. and {Bardeau}, S. and {Sievers}, A. and {Le Petit}, F. and {Le Bourlot}, J. and {Belloche}, A. and {Talbi}, D.},
	doi = {10.1051/0004-6361/201220062},
	eid = {A68},
	eprint = {1210.8178},
	journal = {\aap},
	month = dec,
	pages = {A68},
	primaryclass = {astro-ph.GA},
	title = {{The IRAM-30 m line survey of the Horsehead PDR. II. First detection of the l-C$_{3}$H$^{+}$ hydrocarbon cation}},
	volume = 548,
	year = 2012}

@article{2012ApJ...761...35M,
	adsurl = {http://cdsads.u-strasbg.fr/abs/2012ApJ...761...35M},
	archiveprefix = {arXiv},
	author = {{Micelotta}, E.~R. and {Jones}, A.~P. and {Cami}, J. and {Peeters}, E. and {Bernard-Salas}, J. and {Fanchini}, G.},
	doi = {10.1088/0004-637X/761/1/35},
	eid = {35},
	eprint = {1207.5817},
	journal = {\apj},
	month = dec,
	pages = {35},
	primaryclass = {astro-ph.GA},
	title = {{The Formation of Cosmic Fullerenes from Arophatic Clusters}},
	volume = 761,
	year = 2012}

@article{2012A&A...541A..19A,
	adsurl = {http://cdsads.u-strasbg.fr/abs/2012A%26A...541A..19A},
	archiveprefix = {arXiv},
	author = {{Arab}, H. and {Abergel}, A. and {Habart}, E. and {Bernard-Salas}, J. and {Ayasso}, H. and {Dassas}, K. and {Martin}, P.~G. and {White}, G.~J.},
	doi = {10.1051/0004-6361/201118537},
	eid = {A19},
	eprint = {1202.1624},
	journal = {\aap},
	month = may,
	pages = {A19},
	primaryclass = {astro-ph.GA},
	title = {{Evolution of dust in the Orion Bar with Herschel. I. Radiative transfer modelling}},
	volume = 541,
	year = 2012}

@article{2008A&A...491..797C,
	adsurl = {http://cdsads.u-strasbg.fr/abs/2008A%26A...491..797C},
	archiveprefix = {arXiv},
	author = {{Compi{\`e}gne}, M. and {Abergel}, A. and {Verstraete}, L. and {Habart}, E.},
	doi = {10.1051/0004-6361:200809850},
	eprint = {0809.5026},
	journal = {\aap},
	month = dec,
	pages = {797-807},
	title = {{Dust processing in photodissociation regions. Mid-IR emission modelling}},
	volume = 491,
	year = 2008}

@article{2001PSSAR.186.1521R,
	adsurl = {http://cdsads.u-strasbg.fr/abs/2001PSSAR.186.1521R},
	author = {{Robertson}, J.},
	journal = {Physica Status Solidi Applied Research},
	month = aug,
	pages = {1521-396},
	title = {{Defects in Diamond-Like Carbon}},
	volume = 186,
	year = 2001}

@article{2002MatSciEng..37..129R,
	author = {{Robertson}, J.},
	journal = {Materials Science and Engineering},
	pages = {129-281},
	title = {{Diamond-like amorphous carbon}},
	volume = {37},
	year = {2002}}

@article{2000PhRvB..6114095F,
	adsurl = {http://cdsads.u-strasbg.fr/abs/2000PhRvB..6114095F},
	author = {{Ferrari}, A.~C. and {Robertson}, J.},
	doi = {10.1103/PhysRevB.61.14095},
	journal = {\prb},
	month = may,
	pages = {14095-14107},
	title = {{Interpretation of Raman spectra of disordered and amorphous carbon}},
	volume = 61,
	year = 2000}

@article{2001ApJS..134..263W,
	adsurl = {http://adsabs.harvard.edu/abs/2001ApJS..134..263W},
	author = {{Weingartner}, J.~C. and {Draine}, B.~T.},
	doi = {10.1086/320852},
	eprint = {arXiv:astro-ph/9907251},
	journal = {\apjs},
	month = jun,
	pages = {263-281},
	title = {{Photoelectric Emission from Interstellar Dust: Grain Charging and Gas Heating}},
	volume = 134,
	year = 2001}

@article{2007ApJ...663..320F,
	adsurl = {http://adsabs.harvard.edu/abs/2007ApJ...663..320F},
	archiveprefix = {arXiv},
	author = {{Fitzpatrick}, E.~L. and {Massa}, D.},
	doi = {10.1086/518158},
	eprint = {0705.0154},
	journal = {\apj},
	month = jul,
	pages = {320-341},
	title = {{An Analysis of the Shapes of Interstellar Extinction Curves. V. The IR-through-UV Curve Morphology}},
	volume = 663,
	year = 2007}

@article{2010A&A...510A..36M,
	adsurl = {http://adsabs.harvard.edu/abs/2010A%26A...510A..36M},
	author = {{Micelotta}, E.~R. and {Jones}, A.~P. and {Tielens}, A.~G.~G.~M.},
	doi = {10.1051/0004-6361/200911682},
	journal = {\aap},
	month = feb,
	pages = {A36},
	title = {{Polycyclic aromatic hydrocarbon processing in interstellar shocks}},
	volume = 510,
	year = 2010}

@article{2010A&A...510A..37M,
	adsurl = {http://adsabs.harvard.edu/abs/2010A%26A...510A..37M},
	author = {{Micelotta}, E.~R. and {Jones}, A.~P. and {Tielens}, A.~G.~G.~M.},
	doi = {10.1051/0004-6361/200911683},
	journal = {\aap},
	month = feb,
	pages = {A37},
	title = {{Polycyclic aromatic hydrocarbon processing in a hot gas}},
	volume = 510,
	year = 2010}

@article{1994ApJ...420..307J,
	adsurl = {http://adsabs.harvard.edu/abs/1994ApJ...420..307J},
	author = {{Jochims}, H.~W. and {Ruhl}, E. and {Baumgartel}, H. and {Tobita}, S. and {Leach}, S.},
	doi = {10.1086/173560},
	journal = {\apj},
	month = jan,
	pages = {307-317},
	title = {{Size effects on dissociation rates of polycyclic aromatic hydrocarbon cations: Laboratory studies and astophysical implications}},
	volume = 420,
	year = 1994}

@article{2011A&A...530A..44J,
	adsurl = {http://adsabs.harvard.edu/abs/2011A%26A...530A..44J},
	author = {{Jones}, A.~P. and {Nuth}, J.~A.},
	doi = {10.1051/0004-6361/201014440},
	eid = {A44},
	journal = {\aap},
	month = jun,
	pages = {A44},
	title = {{Dust destruction in the ISM: a re-evaluation of dust lifetimes}},
	volume = 530,
	year = 2011}

@article{2004PhilTransRSocLondA..362.2477F,
	author = {{Ferrari}, A.~C. and {Robertson}, J.},
	journal = {Phil. Trans. R. Soc. Lond. A},
	pages = {2477-2512},
	title = {{Raman spectroscopy of amorphous, nanostructured, diamond-like carbon, and nanodiamond}},
	volume = {362},
	year = {2004}}

@article{2002ApJ...570..697H,
	adsurl = {http://cdsads.u-strasbg.fr/abs/2002ApJ...570..697H},
	author = {{Henry}, R.~C.},
	doi = {10.1086/339623},
	journal = {\apj},
	month = may,
	pages = {697-707},
	title = {{The Local Interstellar Ultraviolet Radiation Field}},
	volume = 570,
	year = 2002}

@article{2005A&A...435..885P,
	adsurl = {http://cdsads.u-strasbg.fr/abs/2005A%26A...435..885P},
	author = {{Pety}, J. and {Teyssier}, D. and {Foss{\'e}}, D. and {Gerin}, M. and {Roueff}, E. and {Abergel}, A. and {Habart}, E. and {Cernicharo}, J.},
	doi = {10.1051/0004-6361:20041170},
	eprint = {arXiv:astro-ph/0501339},
	journal = {\aap},
	month = jun,
	pages = {885-899},
	title = {{Are PAHs precursors of small hydrocarbons in photo-dissociation regions? The Horsehead case}},
	volume = 435,
	year = 2005}

@article{2005A&A...432..895D,
	adsurl = {http://cdsads.u-strasbg.fr/abs/2005A%26A...432..895D},
	author = {{Dartois}, E. and {Mu{\~n}oz Caro}, G.~M. and {Deboffle}, D. and {Montagnac}, G. and {D'Hendecourt}, L.},
	doi = {10.1051/0004-6361:20042094},
	journal = {\aap},
	month = mar,
	pages = {895-908},
	title = {{Ultraviolet photoproduction of ISM dust. Laboratory characterisation and astrophysical relevance}},
	volume = 432,
	year = 2005}

@article{1979PhRvL..42.1151P,
	adsurl = {http://cdsads.u-strasbg.fr/abs/1979PhRvL..42.1151P},
	author = {{Phillips}, J.~C.},
	doi = {10.1103/PhysRevLett.42.1151},
	journal = {Physical Review Letters},
	month = apr,
	pages = {1151-1154},
	title = {{Structure of amorphous (Ge,Si)_{1 - x}Y_{x} alloys}},
	volume = 42,
	year = 1979}

@article{1990MNRAS.247..305J,
	adsurl = {http://cdsads.u-strasbg.fr/abs/1990MNRAS.247..305J},
	author = {{Jones}, A.~P.},
	journal = {\mnras},
	month = nov,
	pages = {305-310},
	title = {{Carbon atom clusters in random covalent networks: PAHs as an integral component of interstellar HAC}},
	volume = 247,
	year = 1990}

@article{1991PSSC..21..199R,
	author = {{Robertson}, J.},
	journal = {Prog. Solid State Chem.},
	pages = {199},
	volume = {21},
	year = {1991}}

@article{1987PhRvB..35.2946R,
	adsurl = {http://cdsads.u-strasbg.fr/abs/1987PhRvB..35.2946R},
	author = {{Robertson}, J. and {O'Reilly}, E.~P.},
	doi = {10.1103/PhysRevB.35.2946},
	journal = {\prb},
	month = feb,
	pages = {2946-2957},
	title = {{Electronic and atomic structure of amorphous carbon}},
	volume = 35,
	year = 1987}

@article{1988PMagL..57..143R,
	adsurl = {http://cdsads.u-strasbg.fr/abs/1988PMagL..57..143R},
	author = {{Robertson}, J.},
	doi = {10.1080/09500838808229624},
	journal = {Philosophical Magazine Letters},
	month = feb,
	pages = {143-148},
	title = {{Clustering and gap states in amorphous carbon}},
	volume = 57,
	year = 1988}

@article{1986AdPhy..35..317R,
	adsurl = {http://cdsads.u-strasbg.fr/abs/1986AdPhy..35..317R},
	author = {{Robertson}, J.},
	doi = {10.1080/00018738600101911},
	journal = {Advances in Physics},
	month = nov,
	pages = {317-374},
	title = {{Amorphous carbon}},
	volume = 35,
	year = 1986}

@article{1980JNS...42...87D,
	author = {{D\"ohler}, G.~H. and {Dandaloff}, R. and {Bilz}, H.},
	journal = {J. Noncryst. Solids},
	pages = {87},
	volume = {42},
	year = {1980}}

@article{2004A&A...423L..33D,
	adsurl = {http://cdsads.u-strasbg.fr/abs/2004A%26A...423L..33D},
	author = {{Dartois}, E. and {Mu{\~n}oz Caro}, G.~M. and {Deboffle}, D. and {d'Hendecourt}, L.},
	doi = {10.1051/0004-6361:200400032},
	journal = {\aap},
	month = aug,
	pages = {L33-L36},
	title = {{Diffuse interstellar medium organic polymers. Photoproduction of the 3.4, 6.85 and 7.25 {$\mu$}m features}},
	volume = 423,
	year = 2004}



\end{document}